\documentclass[preprint,pre,aps]{revtex4}
\usepackage{graphics}
\usepackage{amsmath,amssymb}
\usepackage[dvips]{graphicx}
\begin{document}

\title{Vapor-liquid phase behavior of a size-asymmetric model of ionic fluids confined in a
disordered matrix: the collective variables-based approach}
 \author{O.V. Patsahan, T.M. Patsahan, M.F. Holovko
 }
\affiliation{Institute for Condensed Matter Physics of the National
Academy of Sciences of Ukraine, 1 Svientsitskii St., 79011 Lviv,
Ukraine}

\date{\today}
\begin{abstract}
We develop a theory based on the method of collective variables to study the vapor-liquid equilibrium of
asymmetric ionic fluids confined in a disordered porous matrix.
The approach allows us to formulate the perturbation theory using an extension
of the scaled particle theory  for a description of  a reference system presented as a two-component
hard-sphere  fluid confined in a hard-sphere matrix.
Treating an ionic fluid as a size- and charge-asymmetric primitive model (PM) we derive an explicit expression
for the relevant chemical potential of a confined ionic system which takes into account the third-order correlations
between ions. Using this expression, the phase diagrams for a size-asymmetric PM are calculated  for different matrix
porosities as well as for different sizes of matrix and fluid particles.
It is  observed that  general trends of the coexistence curves with the  matrix porosity are similar to those of simple fluids
under disordered confinement, i.e., the coexistence region gets narrower with a decrease of porosity,
and simultaneously the reduced critical temperature $T_{c}^{*}$ and the critical density $\rho_{i,c}^{*}$ become lower.
At the same time, our results  suggest that an increase in  size asymmetry of oppositely charged ions considerably affects the vapor-liquid
diagrams leading to a faster decrease of $T_{c}^{*}$  and $\rho_{i,c}^{*}$
and even to a disappearance of the phase transition,
especially for the case of small  matrix particles.
\end{abstract}
 \maketitle

 \section{Introduction}
Ionic liquids confined in mesoporous matrices attract a significant attention because
of their specific physicochemical properties and find widespread application in many areas of modern
science and technologies ranging from electrochemistry to biology and medicine.
In particular, such systems are considered as very promising candidates for new types of electrolytes in fuel cells,
supercapacitors, solar cells and batteries. Ionic liquids confined in nanoporous materials are used in catalysis,
sensing and biosensing, gas capture and separation.
Having a high surface area and a large pore volume the biocompatible porous materials are used for a drug delivery,
and ionic liquid encapsulated together with bioactive molecules can be utilized for controlled release of the latter.
Ionic liquids are also considered as efficient porogenic agents used to tune a porous structure (porosity, pore surface area) during
formation of confining material. In the literature one can find a number of excellent reviews devoted to ionic liquids confined
in different porous geometries, in particular,  special attention has been paid to disordered
mesoporous materials \cite{Bideau,Chandra,Zhang2017}.

One of the interesting phenomena related to liquids in confinements, which have been a subject of intensive investigations
for the last decades,  is the vapor-liquid phase transition behavior \cite{Kierlik2001,Gelb,Coasne13}.
It has been shown that a porous medium strongly affects the phase diagram of a guest liquid.
In particular, the phase behavior of liquids in a confined geometry is considerably altered relative to their unconfined (bulk) state.
A distinguished feature of the phase diagrams of such systems as compared with the bulk case is a lowering of both the critical
temperature and the critical density and a narrowing of the coexistence region.
A review of experimental and theoretical efforts in the field is also given in \cite{Gelb,Monson2012,Coasne13}.
However, despite much efforts e being devoted to theoretical studies of these systems, most of them have been related
to studies of simple liquids
or ionic liquids in simple geometries like a single slit or cylindrical pore \cite{Pizio04,Pizio004,Pizio05,Loubet16}.
Moreover, even for simple fluids in disordered confinements
some questions have remained open \cite{Schneider2017}.
At the same time,  the phase behavior of  systems comprised of charged particles confined in disordered matrices has received substantially less attention.
This is mostly related to the fact that it is still a challenge to provide a good quantitative and, in some cases, a qualitative theoretical
description of the phase behavior of ionic fluids even in the bulk. In particular, this concerns the  effects of size and charge asymmetry of ions
on the vapor-liquid phase diagram of ionic fluids with Coulomb-dominated interactions (see \cite{PatMryg12} and references cited therein).
On the other hand, such fluids as ionic liquids are melted organic salts, which are usually characterized
by a significant size asymmetry between cations and anions. Thus, in the present work we develop the method capable of describing 
the thermodynamics of model ionic liquids confined in disorder matrices in order to study the vapor-liquid transition in such systems.

The model most frequently used for ionic fluids is a two-component primitive model (PM) consisting of an electroneutral
mixture of charged hard spheres (HSs)  immersed in a structureless dielectric continuum.
The simulation results for this model have shown that asymmetry in size and charge strongly affects the critical
parameters, i.e.,  the suitably normalized critical temperature decreases with size and charge asymmetry while the critical density
increases with charge asymmetry but decreases with size asymmetry
\cite{Romero-Enrique:00,yan-pablo:01,yan-pablo:02,yan-pablo:02:2,cheong-panagiot:03,kim-fisher-panagiotopoulos:05}.
The original Debye-H\"{u}ckel theory and
the mean spherical approximation (MSA) are not capable of  predicting the trends  observed in simulations
\cite{Gonzalez-Tovar,zuckerman:01}. Moreover,
both  theories predict no dependence on
charge asymmetry in the equisize case.  Progress
in a theoretical description of the effects of asymmetry on the vapor-liquid phase behavior  has been made within the framework of
the associative approach \cite{fisher_aqua_banerjee,QinPrausnitz04}.
However,  certain arbitrariness in determining the association constant is implied
in these theories.
On the other hand, the effect of asymmetry on the vapor-liquid phase diagram has been studied  within the framework of the theory
based on the method of collective variables (CVs)
\cite{patsahan-mryglod-patsahan:06,PatPat09,Patsahan_Patsahan:10,Patsahan_Patsahan:10-2,Patsahan_Patsahan:11}. The theory allows one to take into account the effects of higher-order
correlations between ions and, as a result, to obtain, on an analytical basis, the trends of the critical parameters with charge and size asymmetry
that qualitatively  agree with simulation findings.

It should be noted that the vapor-liquid phase behavior  of  a symmetric ionic model confined in uncharged  pores
of ``simple geometry'',
e.g., of a slit-like or cylindrical
shape  was studied numerically by using the density functional theory \cite{Pizio04,Pizio004,Pizio05}
and quite recently by  the field-theoretical variational approach \cite{Loubet16}.
However, the case of size and charge asymmetry  of ions was not considered in these studies.
On the other hand, the behavior of ionic fluids
in a disordered porous matrix is more complicated because one should take into account the effects of separate pores as well as the effects
of correlations between the ions confined into different pores. Furthermore, a disordered porous matrix is characterized
by specific features as  porosity and  pore surface area. Since the particles composing a matrix are of spherical shape, the mean pore
surface curvature can play a role as well.
To the best of our knowledge, no theoretical results have been obtained until recently for these rather complex
systems.

Fluids confined in disordered porous materials can be treated as  partly-quenched systems
in which some of the degrees of freedom are quenched while the others are annealed \cite {Madden88}.
Partly-quenched systems
containing charges were mainly studied by using
the replica Ornstein-Zernike (ROZ) theory (see  review~\cite{Hribar_Lee:11} and references cited therein). However,
the phase behavior of such systems  has not been considered within this approach.
Moreover, unlike  bulk fluids,
no analytical result has been obtained  from the ROZ theory even for a HS
fluid  confined  in a HS matrix, being the  model of particular importance for the development of perturbation theories.
On the
other hand, based on the  scaled-particle theory (SPT) \cite{Reiss:59}, a pure analytical approach has been   recently  proposed
to describe the thermodynamics of the latter system
\cite{HolDong,PatHol11,HolPat12,HolPat13,Kalyuzhnyi_Holovko_Patsahan_Cummings:14,HolPat15}.
An extension of the SPT developed for a HS fluid confined in a HS matrix,
also referred to as the SPT2 approach \cite{Lebovka-book}, has already been  successfully applied  for the description of reference systems
of different kinds of liquids including the systems  with associative \cite{Kalyuzhnyi_Holovko_Patsahan_Cummings:14} and
anisotropic interactions between particles
\cite{HolShmotPat:14}. More recently this approach has been generalized to the case of a multicomponent HS fluid confined in a multicomponent HS matrix \cite{HolovkoDong16}.

In the previous paper \cite{HolPatPat17-2}, following the idea of Qin and Prausnitz \cite{QinPrausnitz04} for the bulk PM, we developed a
theoretical approach for the study of a vapor-liquid phase transition of a monovalent size-asymmetric PM confined in a disordered
HS matrix. The approach
combines a new extension of the SPT \cite{HolovkoDong16} and the associative
MSA (AMSA)  based on the simplified MSA \cite{QinPrausnitz04}. While the SPT is used for the description of the HS subsystem
presented as a two-component
HS fluid confined in a HS matrix, the
simplified MSA approximates the ionic subsystem by a symmetric ionic fluid  with the effective ion diameter
$\sigma_{+-}=(\sigma_{+}+\sigma_{-})/2$.  However,
the charge asymmetry cannot be taken into consideration within this approach.

In this paper, we continue our systematic studies of the vapor-liquid phase behavior of  ionic systems confined in a disordered porous
medium.
Considering a two-component charge- and size-asymmetric ionic model
confined in a disordered HS matrix, we exploit the idea of  a  partly-quenched model and use the replica trick \cite{Given_Stell:92}.
We extend the work~\cite{HolPatPat16} devoted to a symmetrical ionic model
and develop a CV based theory that allows one  to simultaneously take into account  charge and size asymmetry.
Our approach enables us to formulate the perturbation theory using the SPT for a description of
the thermodynamics of a reference system.
In \cite{HolovkoDong16},
different modifications  derived from the basic SPT formulation are presented and their accuracy is evaluated against the simulation
results. Here, we use the SPT2b approximation which  provides  more accurate results
for excess chemical potentials.

First, we restrict our consideration to the Gaussian approximation and derive the grand  potential of
our partly-quenched system.
For a particular case where
interactions between matrix particles and matrix and fluid particles can be neglected beyond the hard core, we derive an explicit
expression for the relevant chemical potential conjugate to the order parameter in the
approximation that takes into account the effects of correlations up to  third order. To this end, we use the method
proposed in \cite{patsahan-mryglod-patsahan:06,Patsahan_Patsahan:10} for the bulk PM.
Based on this expression, we calculate the
phase diagrams depending on the  characteristics of an ionic fluid and a  HS matrix.  Here, we focus  on  the
size-asymmetric case and consider,  apart from an ion size asymmetry,  different size ratios between the ion particles and the
matrix obstacles.
We analyze how variations in the size-asymmetry parameters and in  the matrix porosity affect the vapor-liquid
phase diagram of a confined ionic model.

The paper is arranged as follows. In section~2, we present a theoretical formalism.
An explicit expression for the  relevant chemical potential which takes into account
the correlation effects beyond the Gaussian approximation  is  derived in this section. Section~3 is devoted to the reference system.
In Section~4, the vapor-liquid phase diagrams of a size-asymmetric ionic fluid confined in HS matrices of different characteristics are presented
and discussed in detail. We conclude in Section~5.

 \section{Model and theory}

\subsection{Model}

We consider a two-component charge- and size-asymmetric ionic model confined in a disordered porous matrix
formed by uncharged  particles. The interaction potentials  between the two matrix particles and  between the ion (cation or anion) 
and the matrix particle
include a short-range attraction or repulsion in addition to a hard-core repulsion. Furthermore,  the ions themselves and the ions
and the matrix particles differ in size.  Therefore, the interactions in the matrix/ionic fluid system can be described by a set
of pairwise interaction potentials: $u_{00}(x)$, $u_{++}(x)$, $u_{--}(x)$, $u_{+-}(x)=u_{-+}(x)$,  $u_{0+}(x)=u_{+0}(x)$, $u_{0-}(x)=u_{-0}(x)$,
where the subscript $0$ refers to the matrix particles and the subscripts $+,-$ refer to the ions.
We assume that the  interaction potentials  between different particles can be split into a reference part  denoted by
index ``r'' and a perturbation part denoted by index ``p''
\begin{eqnarray}
 u_{ij}(x)=u_{ij}^{(r)}(x)+u_{ij}^{(p)}(x),
 \label{split}
 \end{eqnarray}
where $u_{ij}^{(r)}(x)$ is a potential of a short-range repulsion which, generally, describes the mutual impenetrability of the particles,
while  $u_{ij}^{(p)}(x)$ mainly describes the behavior  at moderate and large distances.
The system in which the particles interact via the potentials $u_{ij}^{(r)}(x)$ is regarded as the reference system, $u_{ij}^{(r)}(x)$
is specified in the form of the HS potential. We assume
that the thermodynamic and structural properties of the reference system are known.

We follow the formalism originally proposed in \cite{Madden88}  and  consider  our matrix-fluid
system as a partly-quenched model. This means that our  system contains two subsystems: the first one, the matrix itself,
is composed of particles quenched or frozen in place, while the second subsystem is an annealed  (allowed to equilibrate)
binary ionic fluid which is in equilibrium with the matrix. It is assumed that the matrix particles  were quenched into an
equilibrium configuration corresponding to the Gibbs distribution associated with  a pairwise  interaction potential.
The ionic fluid is treated as a two-component charge- and size-asymmetric PM. In this case, statistical-mechanical
averages used for calculations of thermodynamic properties become double ensemble averages:
the first average is taken over all degrees of freedom of annealed particles  keeping the quenched  particles fixed, and
the other average is performed  over all realizations of a matrix.  To treat the averages   we use the replica method.
It allows us to relate
the matrix averaged quantities to the  thermodynamic quantities of the corresponding fully equilibrated model, referred to as
a replicated model. In our case, the replicated model consists of a matrix and of $s$ identical copies or replicas of the
two-component ionic model.  Each pair of particles has
the same pairwise interaction in this replicated system as in the partly quenched model except that a pair of particles
from different replicas has no interaction. Thus, the interaction potentials between matrix particles, matrix and fluid particles
and fluid and fluid particles can be presented as follows:
\begin{equation}
 u_{00}(r_{0}-r_{0}'), \quad u_{0A}^{\alpha}(r_{0}-r_{\alpha}^{A}), \quad u_{AB}^{\alpha\beta}(r_{\alpha}^{A}-r_{\beta}^{B})\delta_{\alpha\beta},
 \label{int_poten}
\end{equation}
where index ``$0$''refers to matrix particles, Latin indices denote fluid (ion)  species ($A,B=+,-$) and Greek indices denote
replicas ($\alpha,\beta=1,2,\ldots,s$). Furthermore, each interaction potential can be split
into two terms in accordance with (\ref{split}).

\subsection{Collective variables-based approach. Gaussian approximation}

We consider a  $(2s+1)$-component system with the interaction potentials given by (\ref{int_poten}) in the grand canonical ensemble.
Then, using the method of CVs, we can present the equilibrium
grand  partition function of the system  in the form of a functional integral (see \cite{HolPatPat16} and the references therein):
\begin{eqnarray}
\Xi^{\rm{rep}}(s)=\Xi^{\rm{mf}}[\tilde{\nu}_{0},\tilde{\nu}_{A}^{\alpha}]\int ({\rm d}\rho)({\rm d}\omega)
\exp\left[-\frac{\beta}{2}\sum_{{\mathbf k}}\widehat{U}(k)\widehat{\rho}_{{\mathbf k}}\widehat{\rho}_{-{\mathbf k}}\right.
\nonumber\\
\left.
+{\rm i}\sum_{{\mathbf k}}\widehat{\omega}_{{\mathbf k}}\widehat{\rho}_{{\mathbf k}}
+\sum_{n\geq 2}\frac{(-{\rm i})^{n}}{n!}\sum_{{\mathbf{k}}_{1},\ldots,{\mathbf{k}}_{n}}\widehat{{\mathfrak{M}}}_{n}\widehat{\omega}_{\mathbf k_{1}}\widehat{\omega}_{\mathbf k_{2}}\ldots\widehat{\omega}_{\mathbf k_{n}}\delta_{{\bf{k}}_{1}+\ldots +{\bf{k}}_{n}}
\right].
\label{Xi_matrix}
\end{eqnarray}
Here, the following notations are introduced.  $\Xi^{\rm{mf}}$ is the mean-field (MF) part of the grand partition function
 which depends
 on the renormalized partial chemical potentials $\tilde{\nu}_{0}$ and $\tilde{\nu}_{A}^{\alpha}$  [see Eqs.~(\ref{Xi_mf})-(\ref{tilde_nu}) in Appendix~A]; $\beta=1/k_{{\rm B}}T$ with $k_{{\rm B}}$ being the Boltzmann constant,  $T$  the absolute temperature;
$({\rm d}\rho)=({\rm
d}\rho_{0})({\rm d}\rho_{A}^{\alpha})$  ($({\rm d}\omega)=({\rm d}\omega_{0})({\rm d}\omega_{A}^{\alpha})$) denote  volume elements of the phase space of CVs $\rho_{{\mathbf k},0}$
and $\rho_{{\mathbf k},A}^{\alpha}$  ($\omega_{{\mathbf k},0}$ and $\omega_{{\mathbf k},A}^{\alpha}$). CVs $\rho_{{\mathbf k},0}$
and $\rho_{{\mathbf k},A}^{\alpha}$ describe the  fluctuation modes of the  number density of the matrix and fluid species, respectively ($\omega_{{\mathbf k},0}$ and $\omega_{{\mathbf k},A}^{\alpha}$ are
conjugate to $\rho_{{\mathbf k},0}$ and $\rho_{{\mathbf k},A}^{\alpha}$).

$\widehat{U}(k)$ denotes a symmetric $(2s+1)\times(2s+1)$
 matrix of elements:
\begin{eqnarray*}
u_{11}& =&\widetilde{u}_{00}^{(p)}(k)=\widetilde{\varphi}_{00}(k), \nonumber \\
u_{1i}&=&u_{i1} =\widetilde{u}_{0+}^{\alpha(p)}(k)=\widetilde{\varphi}_{0+}(k), \qquad
i\in E, \nonumber \\
u_{1i}&=&u_{i1} =\widetilde{u}_{0-}^{\alpha(p)}(k)=\widetilde{\varphi}_{0-}(k), \qquad
i\in O, \nonumber \\
u_{ii}&=&\widetilde{u}_{++}^{\alpha\alpha(p)}(k)=\widetilde{\varphi}_{++}(k), \qquad
i\in E,  \nonumber \\
u_{ii}&=&\widetilde{u}_{--}^{\alpha\alpha(p)}(k)=\widetilde{\varphi}_{--}(k), \qquad
i\in O,  \nonumber \\
u_{ij}&=&u_{ji}=\widetilde{u}_{+-}^{\alpha\alpha(p)}(k)=\widetilde{\varphi}_{+-}(k), \quad
i\in E, \quad  j=i+1, \nonumber \\
u_{ij}&=&0, \qquad  i\neq j, \quad   j\neq i+1,
\end{eqnarray*}
where the quantities with a ``tilde'' are the Fourier transforms of the corresponding interaction potentials and $E$ ($O$) are even (odd) numbers.
$\widehat{\rho}_{{\mathbf k}}$  indicates a column vector of elements $\rho_{{\mathbf k},0}$, $\rho_{{\mathbf k},+}^{1}$,
$\ldots$, $\rho_{{\mathbf k},+}^{s}$, $\rho_{{\mathbf k},-}^{1}$,
$\ldots$, $\rho_{{\mathbf k},-}^{s}$ and $\widehat{\omega}_{\mathbf k}$ is a row vector of elements $\omega_{{\mathbf k},0}$,
$\omega_{{\mathbf k},+}^{1}$,
$\ldots$, $\omega_{{\mathbf k},+}^{s}$, $\omega_{{\mathbf k},-}^{1}$,
$\ldots$, $\omega_{{\mathbf k},-}^{s}$.

$\widehat{{\mathfrak{M}}}_{n}$ is a  symmetric
$\underbrace{(2s+1)\times(2s+1)\times\ldots\times(2s+1)}_{n}$  matrix   whose elements are cumulants: the $n$th cumulant
coincides with the Fourier transform of the $n$-particle  truncated correlation function \cite{stell} of the reference
system. The  elements of  matrix $\widehat{{\mathfrak{M}}}_{2}$ read as
\begin{eqnarray}
{\mathfrak{M}}_{11}& =&{\mathfrak{M}}_{00}(k), \nonumber \\
{\mathfrak{M}}_{1i}&=&{\mathfrak{M}}_{i1} ={\mathfrak{M}}_{0+}(k), \quad
i\in E,
\qquad
{\mathfrak{M}}_{1i}={\mathfrak{M}}_{i1} ={\mathfrak{M}}_{0-}(k), \quad
 i\in O, \nonumber \\
{\mathfrak{M}}_{ii}&=&{\mathfrak{M}}_{++}^{11}(k), \quad
i\in E,
\qquad
{\mathfrak{M}}_{ii}={\mathfrak{M}}_{--}^{11}(k), \quad
i\in O,  \nonumber \\
{\mathfrak{M}}_{ij}&=&{\mathfrak{M}}_{ji}={\mathfrak{M}}_{+-}^{11}(k), \quad
i\in E, \quad  j\in O,  \quad j=i+1, \nonumber \\
{\mathfrak{M}}_{ij}&=&{\mathfrak{M}}_{ji}={\mathfrak{M}}_{++}^{12}(k), \quad
i,j\in E, \quad i\neq j, \nonumber \\
{\mathfrak{M}}_{ij}&=&{\mathfrak{M}}_{ji}={\mathfrak{M}}_{--}^{12}(k), \quad
i,j\in O, \quad i\neq j, \nonumber \\
{\mathfrak{M}}_{ij}&=&{\mathfrak{M}}_{ji}={\mathfrak{M}}_{+-}^{12}(k), \quad
i\in E, \quad  j\in O, \quad j\neq i+1,
\label{matrix_m2}
\end{eqnarray}
where
\begin{eqnarray}
{\mathfrak{M}}_{00}(k)&=&\overline{\rho_{0}}\delta_{\mathbf{k}}+\overline{\rho_{0}}^{2}\tilde{h}_{00}^{(r)}(k), \quad
{\mathfrak{M}}_{0A}(k)=\overline{\rho_{0}}\,\overline{\rho_{A}}\tilde{h}_{0A}^{(r)}(k),\nonumber \\
{\mathfrak{M}}_{AB}^{\alpha\beta}(k)&=&\overline{\rho^{\alpha}_{A}}\delta_{AB}\delta_{\alpha\beta}\delta_{\mathbf{k}}+
\overline{\rho^{\alpha}_{A}}\,
\overline{\rho^{\beta}_{B}}
\tilde{h}_{AB}^{\alpha\beta(r)}(k),
\label{m_AB}
 \end{eqnarray}
$\overline{\rho_{0}}=\langle N_{0}\rangle_{r}/V$, $\overline{\rho^{\alpha}_{A}}=\langle N_{A}^{\alpha}\rangle_{r}/V$, $\langle\ldots\rangle_{r}$
indicates the  average taken over the reference system and we put $\overline{\rho_{A}^{1}}=\overline{\rho_{A}^{2}}=\ldots=
\overline{\rho_{A}^{s}}=\overline{\rho_{A}}$.
$\tilde{h}_{\ldots}^{\ldots(r)}(k)$ is the Fourier transform of the corresponding pair correlation function of a $(2s+1)$-component 
reference system, $h_{AB}^{11(r)}(r)$ describes the correlations between particles within the same replica, whereas $h_{AB}^{12(r)}(r)$
 describes correlations between the particles from different replicas.
The determinant of the matrix $\widehat{{\mathfrak{M}}}_{2}$ is of the form:
 \begin{eqnarray*}
\det[\widehat{{\mathfrak{M}}}_{2}(s)]&=&\left[({\mathfrak{M}}_{++}^{11}-{\mathfrak{M}}_{++}^{12})({\mathfrak{M}}_{--}^{11}-{\mathfrak{M}}_{--}^{12})
-({\mathfrak{M}}_{+-}^{11}-{\mathfrak{M}}_{+-}^{12})^{2}\right]^{s-1} \nonumber \\
&&
\times
\left\{{\mathfrak{M}}_{00}\left[({\mathfrak{M}}_{++}^{11}+(s-1){\mathfrak{M}}_{++}^{12})({\mathfrak{M}}_{--}^{11}+(s-1){\mathfrak{M}}_{--}^{12})
\right.\right. \nonumber \\
&&
-\left.\left.
({\mathfrak{M}}_{+-}^{11}+(s-1){\mathfrak{M}}_{+-}^{12})^{2}\right]-s{\mathfrak{M}}_{0+}^{2}({\mathfrak{M}}_{--}^{11}+(s-1){\mathfrak{M}}_{--}^{12})\right. \nonumber \\
&&
-
\left.
s{\mathfrak{M}}_{0-}^{2}({\mathfrak{M}}_{++}^{11}
+(s-1){\mathfrak{M}}_{++}^{12})+2s{\mathfrak{M}}_{0+}{\mathfrak{M}}_{0-}({\mathfrak{M}}_{+-}^{11}+(s-1){\mathfrak{M}}_{+-}^{12})
 \right\}.
\end{eqnarray*}

We restrict our consideration to the second order cumulants in Eq.~(\ref{Xi_matrix}). In this case,  after integration we obtain
the grand partition function  of the replicated system  in the Gaussian approximation
\begin{eqnarray}
 \frac{1}{V}\ln\Xi_{\rm{G}}^{\rm{rep}}(s)&=&\frac{1}{V}\ln\Xi^{r}+\frac{\beta}{2}(\overline{\rho_{0}})^{2}\widetilde{\varphi}_{00}(0)
+s\beta\bar\rho_{0}\sum_{A}\overline{\rho_{A}}\widetilde{\varphi}_{0A}(0)
\nonumber \\
&&-\frac{1}{2V}\sum_{{\mathbf{k}}}\ln\left[\det(\widehat{U}\widehat{{\mathfrak{M}}}_{2}+\underline{1})\right],
\label{Ksi_rep_G}
\end{eqnarray}
where $\Xi^{r}$ is the grand partition function of a $(2s+1)$-component reference system.

Using the Legendre transform,  from (\ref{Ksi_rep_G}) one can derive the Helmholtz free energy in the random-phase
approximation (RPA)
\begin{eqnarray*}
\beta f_{\rm{RPA}}(s)&=& \frac{\beta F_{\rm{RPA}}^{\rm{rep}}(s)}{V}=\beta f^{r}-\frac{\rho_{0}^{\rm{rep}}}{2V}
\sum_{{\mathbf{k}}}\beta\widetilde{\varphi}_{00}(k)+\frac{(\rho_{0}^{\rm{rep}})^{2}}{2}\beta\widetilde{\varphi}_{00}(0)
\nonumber \\
&&-\frac{s}{2V}\sum_{A=+,-}\sum_{{\mathbf{k}}}\rho_{A}^{\rm{rep}}\beta\widetilde{\varphi}_{AA}(k)+s\rho_{0}\sum_{A=+,-}
\rho_{A}^{\rm{rep}}\beta\widetilde{\varphi}_{0A}(0)
\nonumber \\
&&+\frac{1}{2V}\sum_{{\mathbf{k}}}\ln\left[\det(\widehat{U}\widehat{{\mathfrak{M}}}_{2}+\underline{1})\right],
\end{eqnarray*}
where $f^{r}$
is the free energy of the reference system,
$\rho_{0}^{\rm{rep}}$ and $\rho_{A}^{\rm{rep}}$  denote the number densities of the matrix  and fluid  particles (cations and anions),
respectively.

Here, we consider a particular case  where interactions between the matrix particles
can be neglected beyond the hard core.
Taking a replica limit of (\ref{Ksi_rep_G})  $-\beta\overline{\Omega}^{G}=\ln\overline{\Xi}^{G}=\lim_{s \to 0}\frac{d}{ds}\ln \Xi^{\rm{\rm{rep}}}_{G}(s)$ we derive,
after some algebra, an expression for the grand  potential of a
partly-quenched system in the Gaussian approximation
\begin{eqnarray}
&&-\beta\overline{\Omega}^{G}=-\beta\overline{\Omega}^{r}+\rho_{0}\sum_{A}\rho_{A}\beta\widetilde{\varphi}_{0A}(0)
-\frac{1}{2}\sum_{{\mathbf{k}}}\ln\left[\det(\widehat{\Phi}_{2}\widehat{{\mathfrak{M}}}_{2}^{c}+\underline{1})\right]
\nonumber \\
&&
-\frac{1}{2}\sum_{{\mathbf{k}}}\frac{1}{\det(\widehat{\Phi}_{2}\widehat{{\mathfrak{M}}}_{2}^{c}
+\underline{1})}\left\{\det(\widehat{\Phi}_{2})\left({\mathfrak{M}}_{++}^{c}{\mathfrak{M}}_{--}^{b}+
{\mathfrak{M}}_{--}^{c}{\mathfrak{M}}_{++}^{b}-2{\mathfrak{M}}_{+-}^{c}{\mathfrak{M}}_{+-}^{b}\right)\right.
\nonumber \\
&&
\left.
+2\beta^{2}\left[\widetilde{\varphi}_{0+}(k)\widetilde{\varphi}_{+-}(k)-\widetilde{\varphi}_{0-}(k)\widetilde{\varphi}_{++}(k)\right]
\left[\overline{\mathfrak{M}}_{0+}{\mathfrak{M}}_{+-}^{c}-\overline{\mathfrak{M}}_{0-}{\mathfrak{M}}_{++}^{c} \right]
\right.
\nonumber \\
&&
\left.
+2\beta^{2}\left[\widetilde{\varphi}_{0+}(k)\widetilde{\varphi}_{--}(k)-\widetilde{\varphi}_{0-}(k)\widetilde{\varphi}_{+-}(k)\right]
\left[\overline{\mathfrak{M}}_{0+}{\mathfrak{M}}_{--}^{c}-\overline{\mathfrak{M}}_{0-}{\mathfrak{M}}_{+-}^{c} \right]
\right.
\nonumber \\
&&
\left.
+\det(\widehat{\Phi}_{3})
\overline{\mathfrak{M}}_{00}\det(\widehat{{\mathfrak{M}}}_{2}^{c})-\overline{\mathfrak{M}}_{00}\sum_{A,B=+,-}\beta\widetilde{\varphi}_{0A}(k){\mathfrak{M}}_{AB}^{c}\beta\widetilde{\varphi}_{0B}(k)
\right.
\nonumber \\
&&
\left.
+2\sum_{A=+.-}\beta\widetilde{\varphi}_{0A}(k)\overline{\mathfrak{M}}_{0A}+\sum_{A,B=+.-}\beta\widetilde{\varphi}_{AB}(k){\mathfrak{M}}_{AB}^{b}
\right\}.
\label{Omega_PQ_0}
\end{eqnarray}
Here, the following notations are introduced.  $\overline{\Omega}^{r}$ is the grand  potential of the reference system consisting of  a two-component HS fluid confined
in a HS matrix, $\rho_{0}=\left.\overline{\rho_{0}}\right\vert_{s=0}$, and $\rho_{A}=\left.\overline{\rho_{A}}\right\vert_{s=0}$.
Matrices $\widehat{\Phi}_{2}$ and $\widehat{\Phi}_{3}$
are of the form:
\begin{equation*}
 \widehat{\Phi}_{2}=
\begin{pmatrix}
  \beta\widetilde{\varphi}_{++}(k) & \beta\widetilde{\varphi}_{+-}(k) \\
  \beta\widetilde{\varphi}_{+-}(k) & \beta\widetilde{\varphi}_{--}(k)
\end{pmatrix},
 \qquad
\widehat{\Phi}_{3}=
\begin{pmatrix}
   0       & \beta\widetilde{\varphi}_{0+}(k) & \beta\widetilde{\varphi}_{0-}(k)\\
\beta\widetilde{\varphi}_{0+}(k) &  \beta\widetilde{\varphi}_{++}(k) & \beta\widetilde{\varphi}_{+-}(k) \\
 \beta\widetilde{\varphi}_{0-}(k)& \beta\widetilde{\varphi}_{+-}(k) & \beta\widetilde{\varphi}_{--}(k)
\end{pmatrix}.
\end{equation*}
${\mathfrak{M}}_{AB}^{c}$ and ${\mathfrak{M}}_{AB}^{b}$ are  elements of the matrices
\begin{equation}
 \widehat{{{\mathfrak{M}}}}_{2}^{c}=
\begin{pmatrix}
  {\mathfrak{M}}_{++}^{c}(k) & {\mathfrak{M}}_{+-}^{c}(k) \\
  {\mathfrak{M}}_{+-}^{c}(k) & {\mathfrak{M}}_{--}^{c}(k) \\
\end{pmatrix}, \qquad
\widehat{{{\mathfrak{M}}}}_{2}^{b}=
\begin{pmatrix}
  {\mathfrak{M}}_{++}^{b}(k) & {\mathfrak{M}}_{+-}^{b}(k) \\
  {\mathfrak{M}}_{+-}^{b}(k) &{\mathfrak{M}} _{--}^{b}(k) \\
\end{pmatrix}.
\label{M-matrix}
\end{equation}
 Superscripts ``c'' and ``b'' in (\ref{M-matrix}) denote  the connected and blocking parts of the cumulants ${\mathfrak{M}}_{AB}$ (or structure factors of the reference system):
\begin{eqnarray}
{\mathfrak{M}} _{AB}(k)={\mathfrak{M}}_{AB}^{c}(k)+{\mathfrak{M}}_{AB}^{b}(k)
=\rho_{A}\delta_{AB} +\rho_{A}\rho_{B}\widetilde{h}_{AB}^{r}(k),
\quad A,B=+,-,
\label{M_AB_s0}
\end{eqnarray}
where
\begin{eqnarray}
 {\mathfrak{M}}_{AB}^{c}(k)&=&\lim_{s \to 0}[{\mathfrak{M}}_{AB}^{11}(k)-{\mathfrak{M}}_{AB}^{12}(k)]=\rho_{A}\delta_{AB}
 +\rho_{A}\rho_{B}\widetilde{h}_{AB}^{r,c}(k), \nonumber \\
 \widetilde{h}_{AB}^{r,c}(k)&=&\lim_{s \to 0}[\widetilde{h}_{AB}^{11(r)}(k)-\widetilde{h}_{AB}^{12(r)}(k)],
 \label{M_ab_c}
\end{eqnarray}
and
\begin{eqnarray}
{\mathfrak{M}}_{AB}^{b}(k)=\lim_{s \to 0}{\mathfrak{M}}_{AB}^{12}=\rho_{A}\rho_{B}\widetilde{h}_{AB}^{r,b}(k), \qquad
\widetilde{h}_{AB}^{r,b}(k)=\lim_{s \to 0}\widetilde{h}_{AB}^{12(r)}(k).
\end{eqnarray}
In Eq.~(\ref{M_AB_s0}),  $\widetilde{h}_{AB}^{r}(k)=\widetilde{h}_{AB}^{r,c}
+\widetilde{h}_{AB}^{r,b}$  is the Fourier transform of the partial pair correlation function with
$\widetilde{h}_{AB}^{r,c(b)}$ being its connected (blocking) part.
 The connected correlation function accounts
for correlations between  a pair of the fluid particles  transmitted  through successive layers of fluid particles while
the blocking correlation function accounts for correlations  between two fluid particles  separated from each other by matrix particles
\cite{Given_Stell:92,Given_Stell:92_2}.

For $\overline{\mathfrak{M}}_{00}$ and $\overline{\mathfrak{M}}_{0A}$, we have
\begin{eqnarray}
\overline{\mathfrak{M}}_{00}(k)&=&\lim_{s \to 0}{\mathfrak{M}}_{00}=\rho_{0}+\rho_{0}^{2}\widetilde{h}_{00}^{r}(k),
 \qquad
 \widetilde{h}_{00}^{r}(k)=\lim_{s \to 0}\widetilde{h}_{00}^{(r)}(k) \nonumber \\
\overline{\mathfrak{M}}_{0A}(k)&=&\lim_{s \to 0}{\mathfrak{M}}_{0A}=\rho_{0}\rho_{A}\widetilde{h}_{0A}^{r}(k),
 \qquad
 \widetilde{h}_{0A}^{r}(k)=\lim_{s \to 0}\widetilde{h}_{0A}^{(r)}(k),
 \label{M00-M0A}
\end{eqnarray}
where $\widetilde{h}_{00}^{r}(k)$ and $\widetilde{h}_{0A}^{r}(k)$ are Fourier transforms of the matrix-matrix and
matrix-fluid correlation functions in a partly-quenched reference system. In Eqs.~(\ref{M_ab_c})-(\ref{M00-M0A}),
${\mathfrak{M}}_{AB}^{\alpha\beta}$, ${\mathfrak{M}}_{00}$, and ${\mathfrak{M}}_{0A}$ are the elements of the
matrix $\widehat{{\mathfrak{M}}}_{2}$ [see Eqs.~(\ref{matrix_m2})-(\ref{m_AB})].

Similarly, one can find the  RPA free energy of a two-component ionic system confined in a disordered porous  matrix. It should be noted that the expression
for free energy of a binary model liquid in a
disordered porous matrix in the RPA  was  derived  in \cite{Kahl:00} in terms of direct correlation functions.

\subsection{Charge- and size-asymmetric primitive model confined in a disordered hard-sphere matrix: Beyond the Gaussian approximation}

We are interested in the vapor-liquid phase diagram of an asymmetric PM  confined in a disordered HS matrix. We assume that
interactions between matrix particles and matrix and fluid particles
can be neglected beyond the hard core. In this case we have
\begin{equation}
\varphi_{00}(r)=0,\quad \varphi_{0+}(r)=0, \quad \varphi_{0-}(r)=0.
\label{PM}
\end{equation}
 The system is electrically neutral: $\sum_{A=+,-}q_{A}\rho_{A}=0$ where $q_{A}$ is a charge of the ion of the
$A$th species, $q_{+}=+zq$, $q_{-}=-q$, $\rho_{A}$  is the number density of the $A$th species.

The model is characterized by the parameters:
\begin{equation}
 \lambda=\frac{\sigma_{+}}{\sigma_{-}}, \qquad z=q_{+}/|q_{-}|
 \label{lambda}
\end{equation}
describing charge and size asymmetry of ions ($\sigma_{A}$ is the diameter of the
$A$th species). In addition, we introduce the parameter $\lambda_{0}$ which describes the size asymmetry between ions 
and matrix particles defined as a size ratio of matrix and negatively charged ions:
\begin{equation}
\lambda_{0}=\sigma_{0}/\sigma_{-}.
\label{lambda0}
\end{equation}

We use the
Weeks-Chandler-Andersen  regularization scheme   for the Coulomb potentials
$\varphi_{AB}(r)$  inside the hard core\cite{wcha}. Then, we have  for $\beta\tilde \varphi_{AB}(k)$:
\begin{eqnarray*}
\beta\tilde\varphi_{++}(k)&=&\frac{4\pi z\sigma_{+-}^{3}}{T^{*}}\frac{(1+\lambda)}{2\lambda}\frac{\sin[2x\lambda/(1+\lambda)]}{x^{3}},
\\
\beta\tilde\varphi_{--}(k)&=&\frac{4\pi\sigma_{+-}^{3} }{T^{*}z}\frac{(1+\lambda)}{2}\frac{\sin[2x/(1+\lambda)]}{x^{3}},
\\
\beta\tilde\varphi_{+-}(k)&=&-\frac{4\pi\sigma_{+-}^{3} }{T^{*}}\frac{\sin(x)}{x^{3}},
\end{eqnarray*}
where   $T^{*}=\frac{k_{B}T\sigma_{+-}}{q^{2}z}$ is the   dimensionless temperature,
$x=k\sigma_{+-}$, and $\sigma_{+-}=(\sigma_{+}+\sigma_{-})/2$.

Our aim here is to derive an analytical expression for the chemical potential conjugate to the order parameter which takes into account the   correlation effects of the order higher  than the second one.
To this end, we follow  a theoretical scheme proposed in \cite{patsahan-mryglod-patsahan:06} for the bulk PM.
We start with the   grand potential of a partly-quenched model  in the Gaussian approximation (\ref{Omega_PQ_0}) under condition
(\ref{PM}) and
we pass from the initial chemical potentials $\nu_{+}$ and $\nu_{-}$ to their linear combinations
\begin{equation*}
 \nu_{1}=\frac{\nu_{+}+z\nu_{-}}{\sqrt{1+z^{2}}}, \qquad
\nu_{2}=\frac{z\nu_{+}-\nu_{-}}{\sqrt{1+z^{2}}}.
\end{equation*}
As was shown in \cite{Patsahan_Patsahan:10}, $\nu_{1}$ is conjugate to the order parameter of the vapor-liquid critical point
\begin{equation}
\xi_{0}=\frac{1}{\sqrt{1+z^2}}\left(\frac{1+z^{2}}{1+z}\rho_{N}+\frac{1-z}{1+z}\rho_{Q}\right),
\label{par_order}
\end{equation}
where $\rho_{N}=\rho_{+}+\rho_{-}$ and $\rho_{Q}=z\rho_{+}-\rho_{-}$ describe long-wavelength
fluctuations of the total number density and charge density, respectively. $\nu_{2}$ is conjugate to $\rho_{Q}$.
 It follows from (\ref{par_order}) that $\xi_{0}\sim \rho_{N}$ for $z=1$.

Then, we present $\nu_{1}$ and $\nu_{2}$ as
\[
\nu_{1}=\nu_{1}^{0}+\varepsilon\Delta\nu_{1}, \qquad
\nu_{2}=\nu_{2}^{0}+\varepsilon\Delta\nu_{2},
\]
where $\nu_{1}^{0}$ and $\nu_{2}^{0}$ are the MF parts of
$\nu_{1}$ and $\nu_{2}$, respectively, and $\Delta\nu_{1}$ and $\Delta\nu_{2}$
are solutions of the equations for chemical potentials.
We self-consistently solve the equations for the relevant chemical potential $\Delta\nu_{1}$  by means of successive approximations.
The procedure of searching for a solution is described in \cite{patsahan-mryglod-patsahan:06,Patsahan_Patsahan:10,PatMryg12}.

The expression for the relevant chemical potential $\nu_{1}$ found in the first nontrivial approximation corresponding to $\nu_{2}=\nu_{2}^{0}$
is of the form
\begin{eqnarray}
\nu_{1}&=&\nu_{1}^{0}+\frac{\sqrt{1+z^{2}}}{2\left[{\mathfrak{M}}_{++}^{c}+2z{\mathfrak{M}}_{+-}^{c}
+z^{2}{\mathfrak{M}}_{--}^{c}\right]}\frac{1}{V}\sum_{{\mathbf
k}}\frac{1}{{\rm
det}\,[\widehat{\Phi}_{2}\widehat{{\mathfrak{M}}}_{2}^{c}+\underline{1}]}\times \nonumber
\\
&& \times\left(\beta\tilde\varphi_{++}(k){\cal F}_{1} \right.\left.
+\beta\tilde\varphi_{--}(k){\cal F}_{2}+2\beta\tilde\varphi_{+-}(k){\cal
F}_{3}\right),
\label{nu1}
\end{eqnarray}
where $\nu_{1}^{0}=\nu_{1}^{r}+\nu_{1}^{se}$ with $\nu_{1}^{r}$ being the combination of the HS chemical potentials
\begin{equation*}
 \nu_{1}^{r}=\frac{\nu_{+}^{r}+z\nu_{-}^{r}}{\sqrt{1+z^{2}}}
\end{equation*}
and $\nu_{1}^{se}$ being the combination of self-energy parts of chemical potentials $\nu_{+}$ and $\nu_{-}$
\begin{equation*}
\nu_{1}^{se}=-\frac{1}{2V\sqrt{1+z^{2}}}\sum_{{\mathbf k}}\left( \beta\tilde\phi_{++}(k)+z\beta\tilde\phi_{--}(k)\right).
\end{equation*}
In addition to the second-order cumulants  ${\mathfrak{M}}_{AB}^{c}$, Eq.~(\ref{nu1}) includes the connected parts of the third order
cumulants ${\mathfrak{M}}_{ABC}^{c}$:
\begin{equation}
{\cal F}_{1}={\mathfrak{M}}_{+++}^{c}+z{\mathfrak{M}}_{++-}^{c}, \qquad {\cal
F}_{2}={\mathfrak{M}}_{+--}^{c}+z{\mathfrak{M}}_{---}^{c},\qquad {\cal
F}_{3}={\mathfrak{M}}_{++-}^{c}+z{\mathfrak{M}}_{+--}^{c}.
\label{F-3}
\end{equation}
In (\ref{nu1}), ${\mathfrak{M}}_{AB}^{c}$ and ${\mathfrak{M}}_{ABC}^{c}$ are approximated by their values in the long-wavelength limit.
A general form of  Eq.~(\ref{nu1}) is similar to that for the bulk case obtained in \cite{Patsahan_Patsahan:10}. However, the main
difference  concerns the reference system.
Below, we  consider the reference system in more detail.

\section{Reference system: thermodynamic properties from the scaled particle theory}
We start with   general  relationships valid for a multicomponent system, in particular,
a  recurrent formula allowing us to derive the third order cumulants in the long-wavelength limit \cite{stat}
\begin{equation}
{\mathfrak{M}}_{\alpha_{1}\alpha_{2}\ldots\alpha_{n}}={\mathfrak{M}}_{\alpha_{1}\alpha_{2}\ldots\alpha_{n}}(0,\ldots)=\frac{\partial
{\mathfrak{M}}_{\alpha_{1}\alpha_{2}\ldots\alpha_{n-1}}(0,\ldots)}{\partial\nu_{\alpha_{n}}}
\label{nth-cumulant}
\end{equation}
and  the equation given by Kirkwood and Buff which relates the thermodynamic properties  with
the partial structure factors at $k=0$ \cite{kirkbuf}
\begin{equation}
S_{\alpha_{1}\alpha_{2}}(0)=\frac{1}{\sqrt{\rho_{\alpha_{1}}\rho_{\alpha_{2}}}}{\mathfrak{M}}_{\alpha_{1}\alpha_{2}}(0)=
\frac{1}{\sqrt{\rho_{\alpha_{1}}\rho_{\alpha_{2}}}}\frac{\lvert A\rvert_{\alpha_{1}\alpha_{2}}}{\det(A)}.
\label{A_12}
\end{equation}
In Eq.~(\ref{A_12}),  $A$
is a matrix  with elements given by
$A_{\alpha_{1}\alpha_{2}}=\left(\partial\nu_{\alpha_{1}}/{\partial\rho}_{\alpha_{2}}\right)_{T,\rho_{\alpha_{3}}}$,
$\lvert A\rvert_{\alpha_{1}\alpha_{2}}$ indicates the cofactor of the elements
$A_{\alpha_{1}\alpha_{2}}$.
Using the Ornstein-Zernike equation we obtain from (\ref{A_12})
\begin{equation}
\left(\frac{\partial\nu_{\alpha_{1}}}{\partial\rho_{\alpha_{2}}}\right)_{T,\rho_{\alpha_{3}}}=\frac{\delta_{\alpha_{1}\alpha_{2}}}
{\rho_{\alpha_{1}}}-\tilde{c}_{\alpha_{1}\alpha_{2}}(0),
 \label{C_12}
\end{equation}
where $\tilde{c}_{\alpha_{1}\alpha_{2}}(0)$ is the Fourier transform of the partial direct correlation functions \cite{hansen_mcdonald}
at $k=0$.
In  \cite{Kahl:00}, general expressions  were presented for thermodynamic quantities and relations for a two-component system confined
in a disordered matrix. In particular, it was shown that in this case,
Eqs.~(\ref{C_12}) are satisfied
for the connected parts of $\tilde{c}_{\alpha_{1}\alpha_{2}}(0)$ [see Eqs.~(47)-(49) in \cite{Kahl:00}].
Using (\ref{nth-cumulant}),  ${\mathfrak{M}}_{\alpha_{1}\alpha_{2}\alpha_{3}}(0,0)$ can be expressed
in terms of the partial structures factors $S_{\alpha_{1}\alpha_{2}}(0)$ and their derivatives.  The corresponding formulas for
a two-component
system are given in Appendix~B. The same formulas hold for the connected parts of the  quantities entering
Eqs.~(\ref{M_+++}) and (\ref{M_++-}).

Now, we turn back to our reference system which consists of a two-component HS fluid confined in a one-component HS  matrix.
The matrix is characterized by  HS obstacles of diameter $\sigma_{0}$ and different types of porosity, namely,   geometrical
porosity $\phi_{0}$ and two probe-particle porosities  $\phi_{+}$ and $\phi_{-}$ for the two fluid species. The probe-particle
porosity $\phi_{+}$ ($\phi_{-}$) is defined by the excess
value of the chemical potential of a fluid particle with  diameter $\sigma_{+}$ ($\sigma_{-}$)
in the limit of infinite dilution and, hence, takes into account  the size of adsorbate species \cite{HolovkoDong16}.
The geometrical porosity
$\phi_{0}$ is independent of  adsorbate. It defines  a ''bare'' pore volume of the matrix and  can be
considered as a more general characteristic.  For the HS matrix
$\phi_{0}=1-\eta_{0}$,
where $\eta_{0}=\pi\rho_{0}\sigma_{0}^{3}/6$, $\rho_{0}=N_{0}/V$ is the
number density of  matrix particles.

Using the results
obtained for an $n$-component HS fluid in an $m$-component
HS matrix \cite{HolovkoDong16}  we  find analytical expressions for the chemical potentials $\nu_{+}^{r}$ and $\nu_{-}^{r}$.
The expression for $\nu_{+}^{r}$ in the SPT2b approximation providing the best accuracy   reads
\begin{eqnarray}
\nu_{+}^{r}&=&\nu_{+}^{SPT2b}=\ln(\Lambda_{+}^{3}\eta_{+})-\ln(\phi_{+})+k_{+}^{1}\frac{\eta_{i}/\phi_{0}}{1-\eta_{i}/\phi_{0}}+k_{+}^{2}
\left(\frac{\eta_{i}/\phi_{0}}{1-\eta_{i}/\phi_{0}}\right)^2
 \nonumber \\
&&+k_{+}^{3}\left(\frac{\eta_{i}/\phi_{0}}{1-\eta_{i}/\phi_{0}}\right)^3-\ln\left(1-\frac{\eta_{i}}{\phi}\right)\left\{1
-\frac{\phi}{\eta_{i}}\left[1-\frac{\phi}{\phi_{+}}\frac{(\eta_{+}+\lambda^{3}\eta_{-})}{\eta_{i}}\right]\right\}
\nonumber \\
&&
-\frac{\phi_{0}}{\eta_{i}}\ln\left(1-\frac{\eta_{i}}{\phi_{0}}\right)
\left(1-\frac{\eta_{+}
+\lambda^{3}\eta_{-}}{\eta_{i}}\right)
+\frac{(\eta_{+}+\lambda^{3}\eta_{-})}{\eta_{i}}\left(1-\frac{\phi}{\phi_{+}}\right),
\label{nu_+_hs}
\end{eqnarray}
where $\eta_{i}=\eta_{+}+\eta_{-}$, $\eta_{A}=\frac{\pi}{6}\rho_{A}\sigma_{A}^{3}$ ($A=+,-$), and
\begin{equation}
\phi^{-1}=\frac{1}{\eta_{i}}\left(\frac{\eta_{+}}{\phi_{+}}+\frac{\eta_{-}}{\phi_{-}}\right).
\label{phi}
\end{equation}
For  $\phi_{+}$, we have
\begin{eqnarray}
&&\phi_{+}=(1-\eta_{0})\exp\left\{-\frac{6\tau\lambda}{(1+\lambda)}\frac{\eta_{0}}{(1-\eta_{0})}
\left[1+\frac{\tau\lambda}{1+\lambda}
\left(2+\frac{3\eta_{0}}{1-\eta_{0}}\right)
\right.
\right.\nonumber\\
&&
\left.\left.
+\frac{4}{3}\left(\frac{\tau\lambda}{1+\lambda}\right)^{2}
\left(1+\frac{3\eta_{0}}{1-\eta_{0}}
+3\left(\frac{\eta_{0}}
{1-\eta_{0}}\right)^{2}\right)\right]\right\},
\label{phi_+_HS}
\end{eqnarray}
where the parameter
\begin{equation}
\tau=\frac{\sigma_{+-}}{\sigma_{0}}=\frac{1+\lambda}{2\lambda_{0}}
\label{tau}
\end{equation}
is introduced.
$\phi_{-}$ is obtained from (\ref{phi_+_HS}) by replacing $\lambda$ with $1/\lambda$.

The expressions for coefficients $k_{+}^{i}$ are as follows:
\begin{eqnarray}
 k_{+}^{1}&=&\frac{\eta_{-}\lambda(\lambda^{2}+3\lambda+3)+7\eta_{+}}{\eta_{i}}
+\frac{3\tau\lambda\eta_{0}\left[\eta_{-}(\lambda^2+6\lambda+1)+8\eta_{+}\right]}{\eta_{i}(1+\lambda)(1-\eta_{0})}\nonumber \\
&&
+\frac{6\tau^2\lambda^2\eta_{0}(1+2\eta_{0})\left[\eta_{-}(1+\lambda)+2\eta_{+}\right]}{\eta_{i}(1+\lambda)^2(1-\eta_{0})^2},
\label{k+1}
\end{eqnarray}
\begin{eqnarray}
k_{+}^{2}&=&\frac{3}{2}\frac{\left[\eta_{-}^2\lambda^2(2\lambda+3)+2\eta_{+}\eta_{-}\lambda(\lambda+4)+5\eta_{+}^2\right]}{\eta_{i}^2}
\nonumber \\
&&
+\frac{3\tau\lambda\eta_{0}\left[2\eta_{-}^{2}\lambda(2+3\lambda)+\eta_{+}\eta_{-}(\lambda^2+14\lambda+5)+10\eta_{+}^{2}\right]}
{\eta_{i}^2(1+\lambda)(1-\eta_{0})}
\nonumber \\
&&
+\frac{6\tau^2\lambda^2\eta_{0}\left[\eta_{-}(4\eta_{0}\lambda+\eta_{0}+\lambda)+\eta_{+}(5\eta_{0}+1)\right]}{(1+\lambda)^2
(1-\eta_{0})^2\eta_{i}},
\label{k+2}
\end{eqnarray}
\begin{eqnarray}
 k_{+}^{3}=3\left[\frac{2\tau\lambda\eta_{0}}{(1+\lambda)(1-\eta_{0})}+\frac{\eta_{+}
 +\lambda\eta_{-}}{\eta_{i}}\right]^2\frac{\eta_{+}+\lambda
 \eta_{-}}{\eta_{i}}.
 \label{k+3}
\end{eqnarray}
The expression for $\nu_{-}^{r}$ can be obtained from Eqs.~(\ref{nu_+_hs})-(\ref{k+3}) by replacing $\eta_{+}$ with $\eta_{-}$  and vice versa
as well as by replacing $\lambda$ with $1/\lambda$, $\phi_{+}$ with $\phi_{-}$, and $k_{+}^{i}$ with $k_{-}^{i}$ ($i=1,2,3$).

Based on the equations for $\nu_{+}^{r}$ and $\nu_{-}^{r}$ obtained above one can derive analytic expressions  for
${\mathfrak{M}}_{AB}^{c}(0)$ and ${\mathfrak{M}}_{ABC}^{c}(0,0)$  which enter  the equation
(\ref{nu1}) for the relevant chemical potential. In particular,  for  $S_{AB}^{c}={\mathfrak{M}}_{AB}^{c}(0)/\sqrt{\rho_{A}\rho_{B}}$ we have
\begin{eqnarray}
 S_{++}^{c}&=&\left(\frac{\partial\nu_{-}^{r}}{\partial \eta_{-}}\right)_{\eta_{+}}\frac{\eta_{-}}{\det(A_{2}^{r})}, \qquad
 S_{--}^{c}=\left(\frac{\partial\nu_{+}^{r}}{\partial \eta_{+}}\right)_{\eta_{-}}\frac{\eta_{+}}{\det(A_{2}^{r})}, \nonumber \\
 S_{+-}^{c}&=&-\left(\frac{\partial\nu_{-}^{r}}{\partial \eta_{+}}\right)_{\eta_{-}}\frac{\sqrt{\lambda^{3}\eta_{+}\eta_{-}}}{\det(A_{2}^{r})},
\label{S_ij}
\end{eqnarray}
where
\begin{equation}
\det(A_{2}^{r})=\eta_{+}\eta_{-}\left[\left(\frac{\partial\nu_{+}^{r}}{\partial\eta_{+}}\right)_{\eta_{-}}
\left(\frac{\partial\nu_{-}^{r}}{\partial\eta_{-}}\right)_{\eta_{+}}
 -\left(\frac{\partial\nu_{+}^{r}}{\partial \eta_{-}}\right)_{\eta_{+}}\left(\frac{\partial\nu_{-}^{r}}{\partial \eta_{+}}\right)_{\eta_{-}}\right].
 \label{M_ij}
\end{equation}
Taking into account (\ref{F-3}) and formulas (\ref{M_+++})-(\ref{M_++-}) from Appendix~B, the coefficients 
${\cal F}_{1}$,  ${\cal F}_{2}$, and ${\cal F}_{3}$ can be written in the form:
\begin{eqnarray}
\frac{{\cal F}_{1}}{\rho_{+}}&=&\left(S_{++}^{c}+\eta_{+}\frac{\partial S_{++}^{c}}{\partial\eta_{+}}\right)\left(S_{++}^{c}+z\sqrt{z}S_{+-}^{c}\right)+\frac{\eta_{-}}{\sqrt{z}}
\frac{\partial S_{++}^{c}}{\partial\eta_{-}}\left(S_{+-}^{c}+z\sqrt{z}S_{--}^{c}\right), \nonumber \\
\frac{{\cal F}_{2}}{\rho_{-}}&=&\frac{1}{\sqrt{z}}\left(S_{--}^{c}+\eta_{-}\frac{\partial S_{--}^{c}}{\partial\eta_{-}}\right)\left(S_{+-}^{c}+z\sqrt{z}S_{--}^{c}\right)+
\eta_{+}\frac{\partial S_{--}^{c}}{\partial\eta_{+}}\left(S_{++}^{c}+z\sqrt{z}S_{+-}^{c}\right), \nonumber\\
\frac{{\cal F}_{3}}{\sqrt{\rho_{+}\rho_{-}}}&=&S_{+-}^{c}\left(S_{++}^{c}+\eta_{+}\frac{\partial S_{++}^{c}}{\partial\eta_{+}}\right)+zS_{+-}^{c}\left(S_{--}^{c}+\eta_{-}\frac{\partial S_{--}^{c}}
{\partial\eta_{-}}\right)+\frac{1}{\sqrt{z}}S_{--}^{c}\eta_{-}\frac{\partial S_{++}^{c}}{\partial\eta_{-}}\nonumber \\
&&
+z\sqrt{z}S_{++}^{c}\eta_{+}\frac{\partial S_{--}^{c}}{\partial\eta_{+}}.
\label{bar_Fij}
\end{eqnarray}
Explicit expressions for (\ref{S_ij})-(\ref{bar_Fij}) are too long to be presented here.

\section{Results and discussion}
Using the equations (\ref{nu1})-(\ref{F-3})  (as well as the expressions from Sec.~3), we study the vapor-liquid phase diagrams 
of the PM confined in a HS matrix. Here, we focus on a monovalent size-asymmetric PM. Because of symmetry with
respect to the exchange of ``$+$'' and ``$-$'' ions, only $\lambda >1$ or $\lambda <1$ need be considered in this case. 
Supplementing the above-mentioned equations by the Maxwell construction, we calculate the coexistence curves  and the
corresponding critical parameters for
different values of  size ratios $\lambda$ and $\lambda_{0}$  [see (\ref{lambda})-(\ref{lambda0})] and  for different matrix porosities
$\phi_{0}$. Estimates of the critical temperature and the critical density are given by their values for which the maxima and minima of the van der Waals loops
coalesce. The reduced temperature and the reduced density are chosen in the conventional form,
which is common to the works dealing with the phase behavior of an asymmetric PM in the bulk state (see, for example, Ref.~ \cite{Romero-Enrique:00})
\begin{equation}
T^{*}=\frac{k_{\mathrm{B}}T\sigma_{+-}}{q^{2}z}, \qquad
\rho_{i}^{*}=\rho_{i}\sigma_{+-}^{3},
\label{crit_par}
\end{equation}
 where $\rho_{i}=\rho_{+}+\rho_{-}$ is the total ionic number density. It should be  noted that  for the bulk PM our theory predicts a reduction of the
coexistence regions as well as a decrease of the critical
parameters $T_{c}^{*}$ and $\rho_{i,c}^{*}$ with an increase of size
asymmetry \cite{Patsahan_Patsahan:10,Patsahan_Patsahan:10-2}. This behavior qualitatively agrees with simulation results \cite{Romero-Enrique:00,yan-pablo:01,yan-pablo:02,yan-pablo:02:2,cheong-panagiot:03,kim-fisher-panagiotopoulos:05}.
\begin{figure}[htb]
\begin{center}
\includegraphics[clip,width=0.32\textwidth,
angle=0]{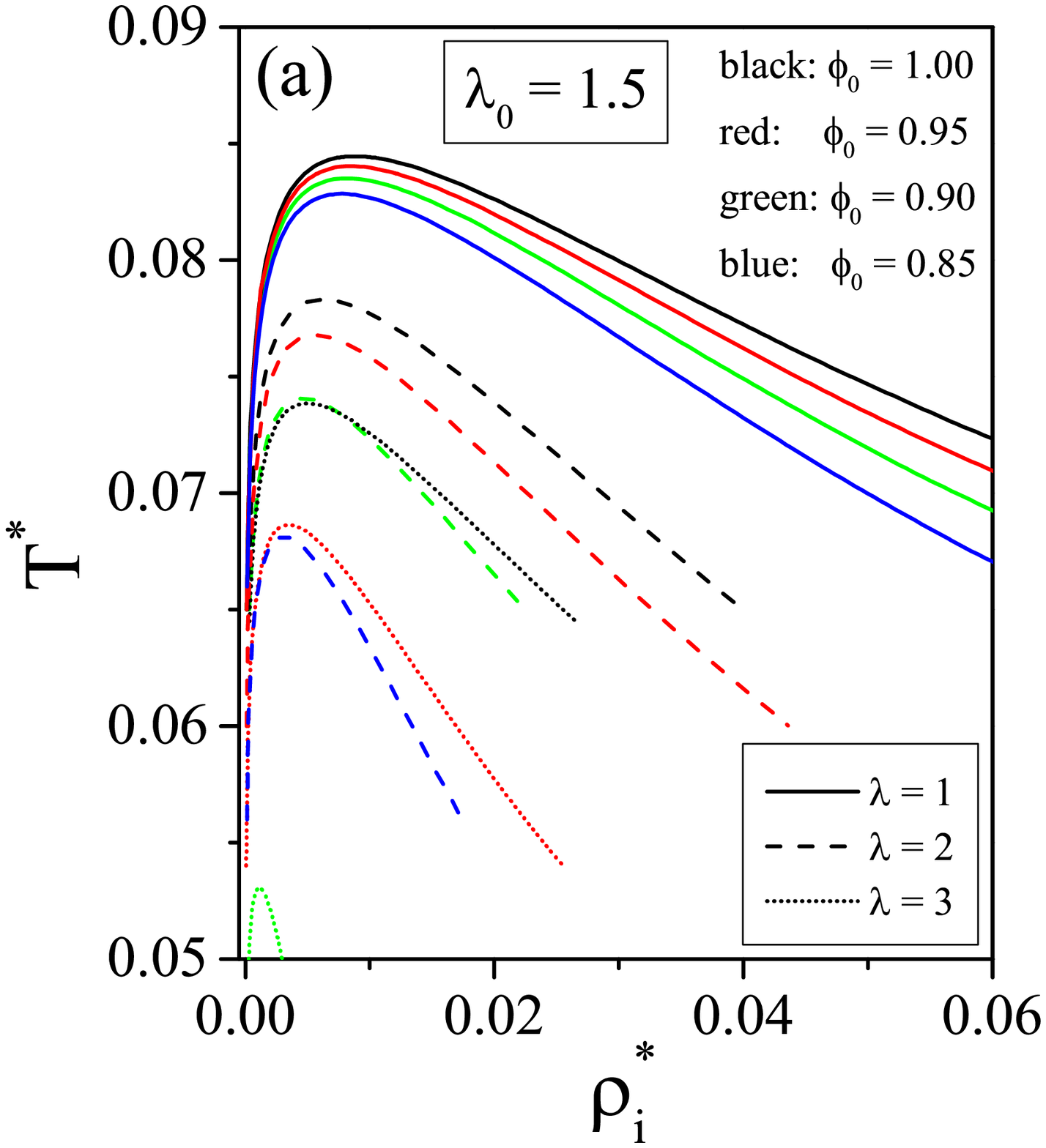}
\includegraphics[clip,width=0.32\textwidth,
angle=0]{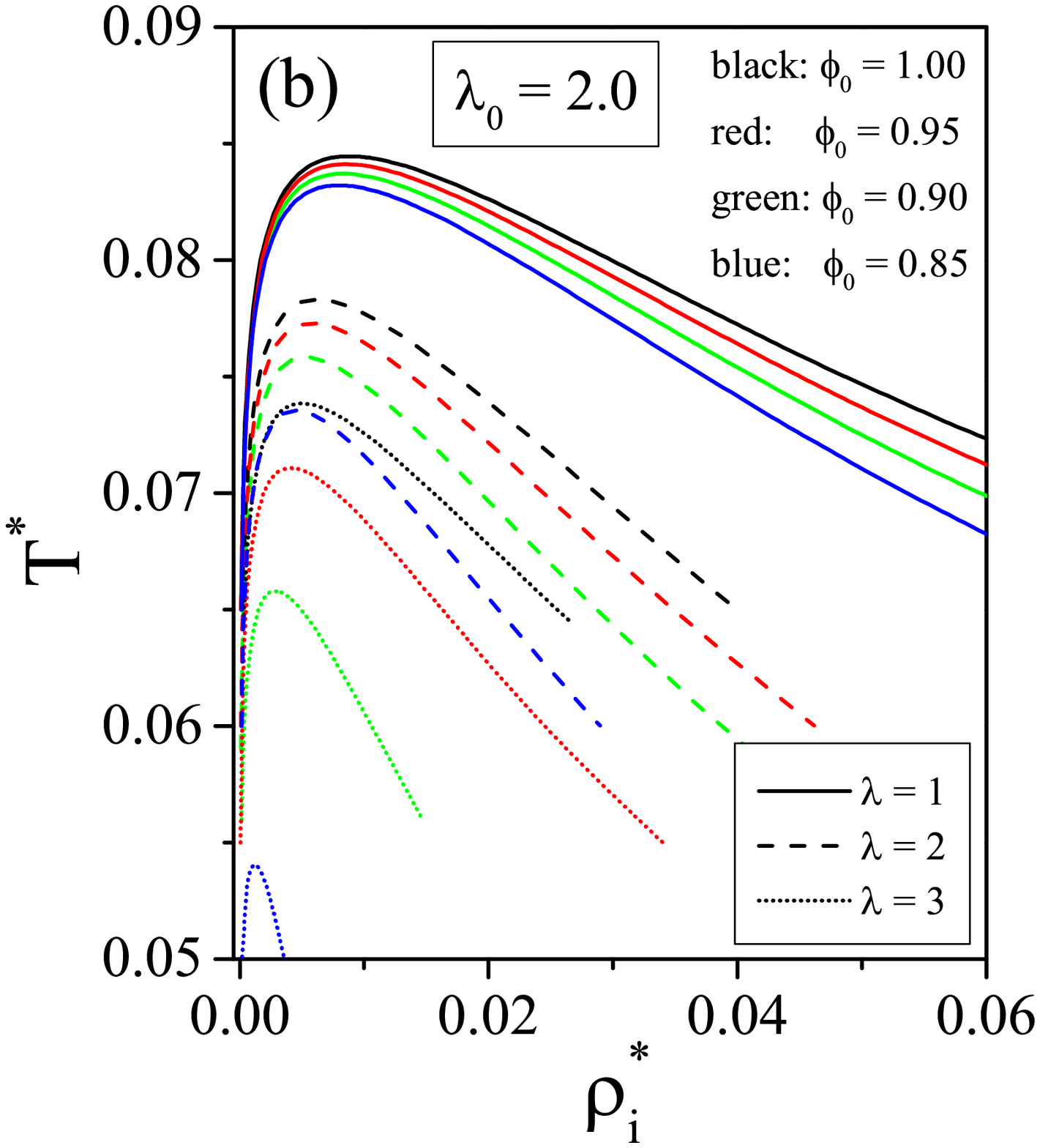}
\includegraphics[clip,width=0.32\textwidth,
angle=0]{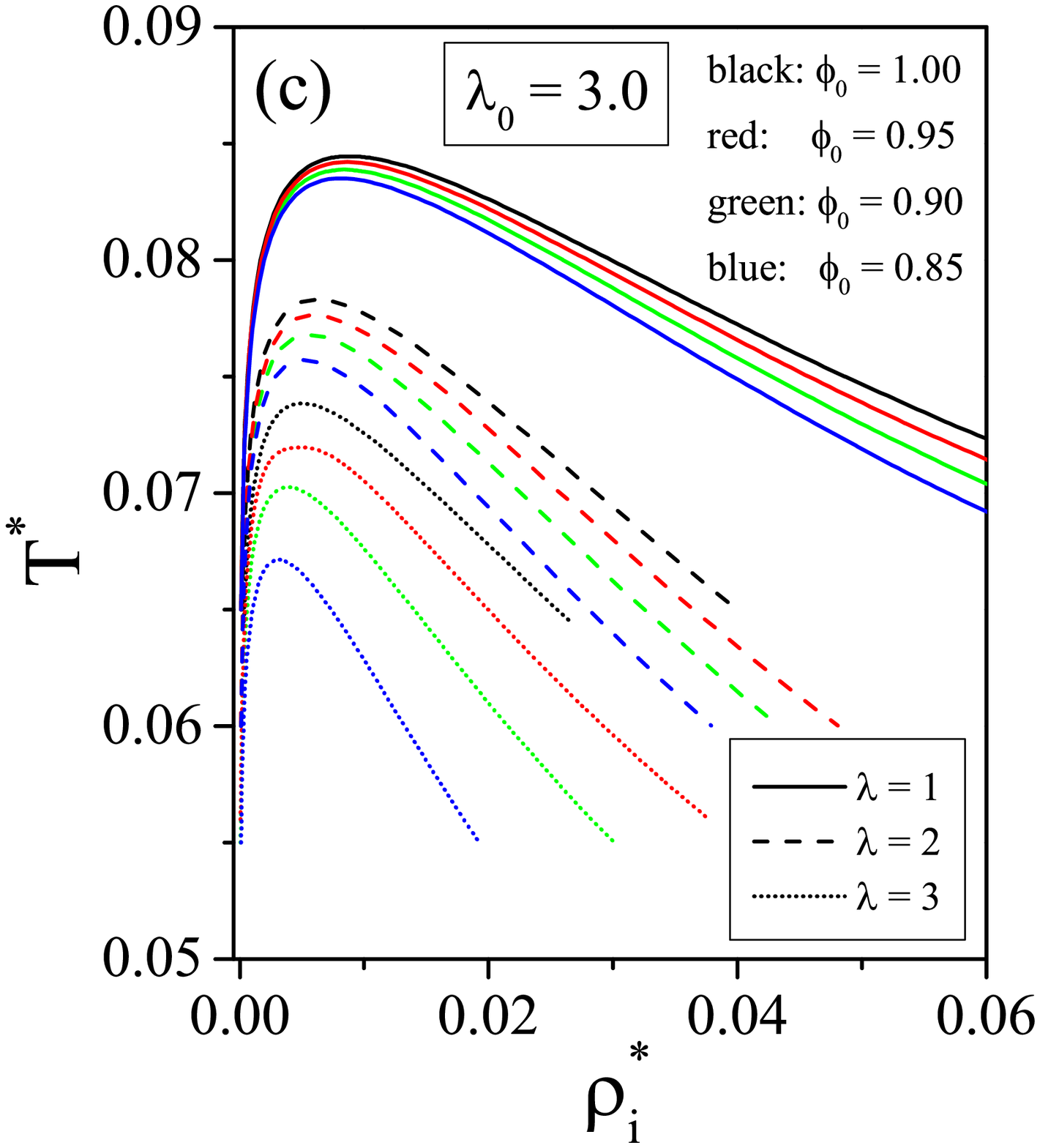} \caption{\label{fig:phase}
(Color online) Vapor-liquid phase diagrams at the fixed size ratios $\lambda_{0}=1.5$ (a),  $\lambda_{0}=2.0$ (b), 
and $\lambda_{0}=3.0$~(c) [see Eq.~(\ref{lambda0})]. 
In each case, data are shown for different ratios of ion size asymmetry $\lambda$ 
as indicated in the legends and for different matrix porosities $\phi_{0}$. At the fixed $\lambda$, $\phi_{0}=1$, $0.95$, $0.9$, and $0.85$ 
(from top to bottom). The bulk case ($\phi_{0}=1$) is presented for comparison. 
Temperature $T^{*}$  and the ion density $\rho_{i}^{*}$ are in dimensional reduced units defined in Eqs.~(\ref{crit_par}).}
\end{center}
\end{figure}

The calculated phase diagrams in the ($T^{*}$-$\rho_{i}^{*}$) plane for $\lambda_{0}=1.5$, $2$ and $3$ are shown 
in Figs.~\ref{fig:phase}(a)--\ref{fig:phase}(c). In each figure, for the given $\lambda_{0}$, we show the coexistence curves 
for $\lambda=1$, $2$ and $3$ and for the three values of matrix porosity $\phi_{0}=0.85$, $0.9$ and $0.95$. 
The bulk case, $\phi_{0}=1.0$, is shown for comparison. For the fixed $\lambda_{0}$ and $\lambda$, the phase diagrams 
demonstrate the usual behavior of confined fluids, i.e., both the critical temperature and the critical density decrease when the porosity decreases and simultaneously the coexistence region becomes narrower. An increase of  size asymmetry of the ions which corresponds to an increase of the parameter $\lambda$ essentially strengthens the tendency of $T_{c}^{*}$ and $\rho_{i,c}^{*}$ towards lower values.
By contrast, an increase of the size of matrix particles, i.e., an increase of the parameter $\lambda_{0}$,
leads to the opposite effect. Hence, in Figs.~\ref{fig:phase}(a)--\ref{fig:phase}(c) one can observe a competition between different effects controlled by the parameters $\phi_0$, $\lambda$ and $\lambda_{0}$.
Since the matrix is totally defined by its porosity and by the size of matrix particles $\sigma_{0}$,
the parameters $\phi_0$ and $\lambda_{0}=\sigma_{0}/\sigma_{-}$ are responsible for the confinement effects.
It is seen from  Fig.~\ref{fig:phase} that the strongest confinement effect is obtained
for the lowest porosity and for the smallest size of matrix particles, i.e., for $\phi_0=0.85$ and $\lambda_{0}=1.5$ [see Fig.~\ref{fig:phase}(a)].
In this case, the critical parameters $T_{c}^{*}$ and $\rho_{i,c}^{*}$  dramatically decrease with an increase
of the ion size asymmetry  $\lambda$. Moreover, for $\lambda=3$, the critical temperature $T_{c}^{*}$ is so low that the vapor-liquid
phase transition gets beyond  the temperature range considered in our study.
On the other hand, the phase transition has been obtained for larger matrix particles [see Figs.~\ref{fig:phase}(b) and \ref{fig:phase}(c)], although for $\lambda_{0}=2$, the coexistence region  is rather small.

\begin{figure}[htb]
\begin{center}
\includegraphics[clip,width=0.32\textwidth,
angle=0]{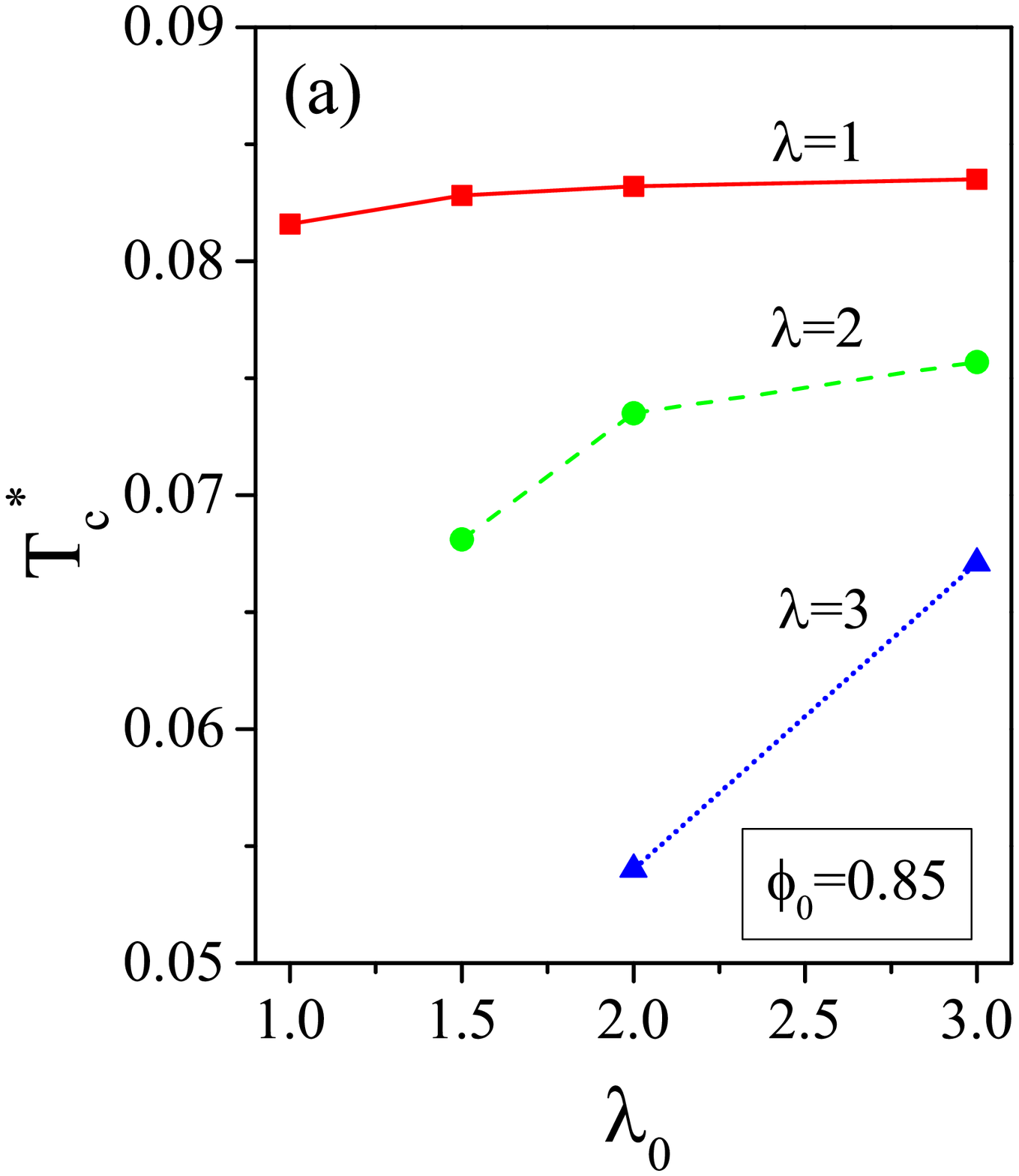}
\includegraphics[clip,width=0.32\textwidth,
angle=0]{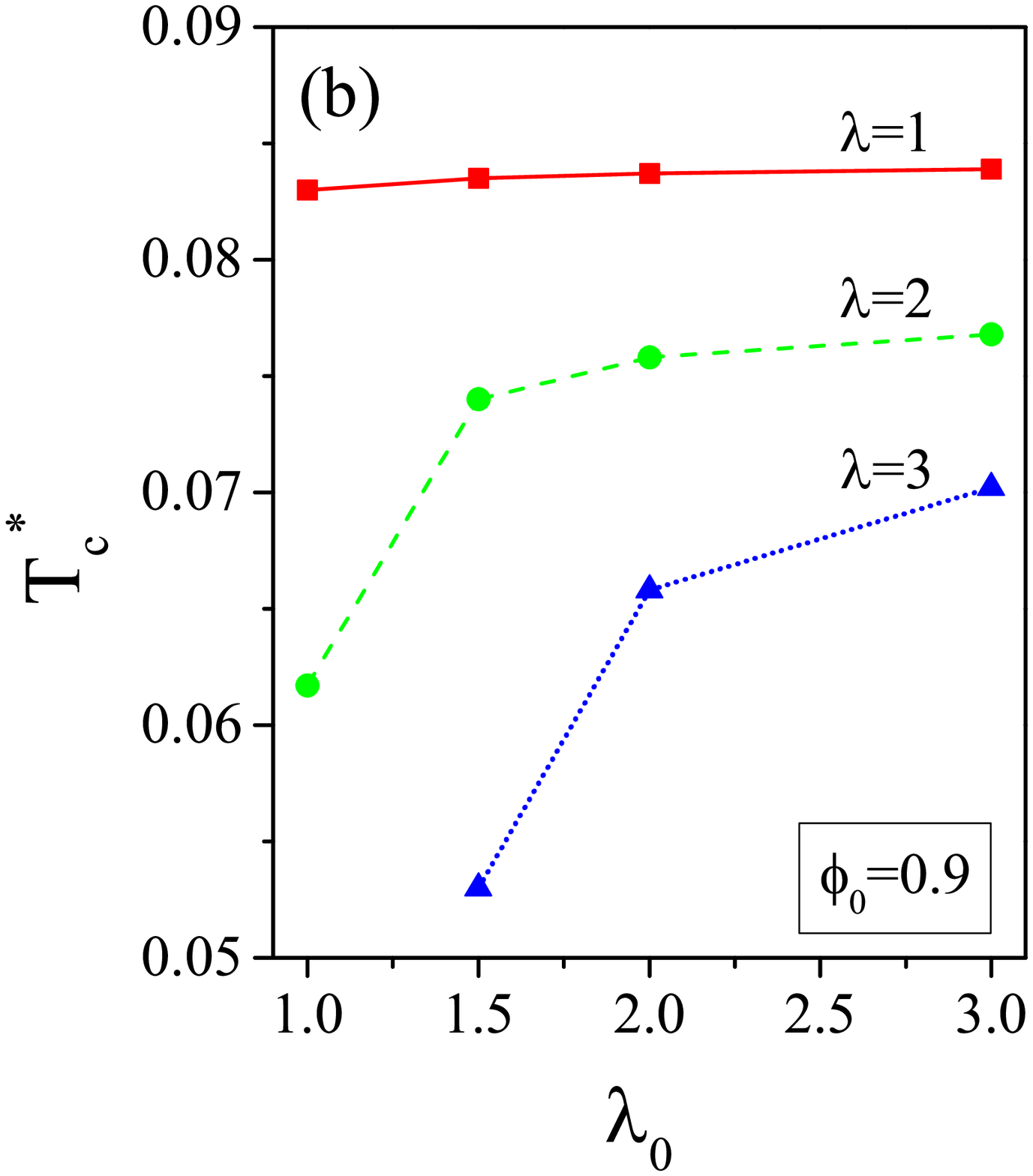}
\includegraphics[clip,width=0.32\textwidth,
angle=0]{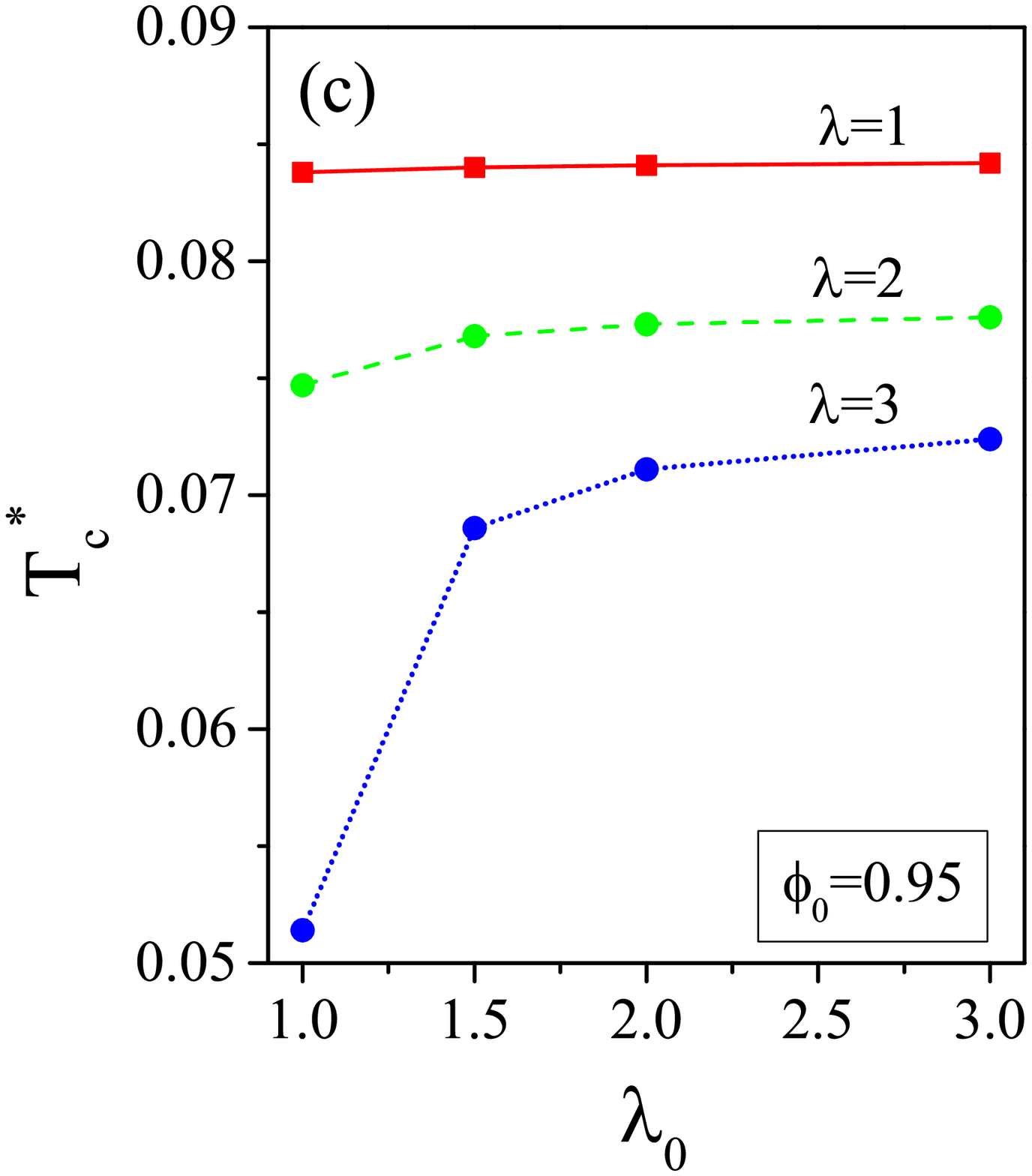} \caption{\label{fig:Tclamb0}
(Color online) Dependence of the critical temperature $T_{c}^{*}$ on the size of matrix particles $\lambda_{0}$ [see Eq.~(\ref{lambda0})]
for different  matrix porosities: (a) $\phi_{0}=0.85$, (b) $\phi_{0}=0.9$, and (c) $\phi_{0}=0.95$. In each case, data are shown for different values of the ion size ratio $\lambda$.
}
\end{center}
\end{figure}

In Figs.~\ref{fig:Tclamb0}(a)--\ref{fig:Tclamb0}(c), we present the dependence of the critical 
temperature $T_{c}^{*}$ on $\lambda_{0}$ in more detail.  It is seen, that for a symmetric ionic fluid ($\lambda=1$), the dependence of $T_{c}^{*}$ on $\lambda_{0}$ is weak, especially for large porosity ($\phi_{0}=0.95$).
For an asymmetric ionic fluid ($\lambda=2$ and $3$), this dependence drastically changes,
i.e., $T_{c}^{*}$ starts to sharply decrease  at  small values of $\lambda_{0}$. This trend is more prominent
when the porosity $\phi_{0}$ is lower.

\begin{figure}[htb]
\begin{center}
\includegraphics[clip,width=0.32\textwidth,
angle=0]{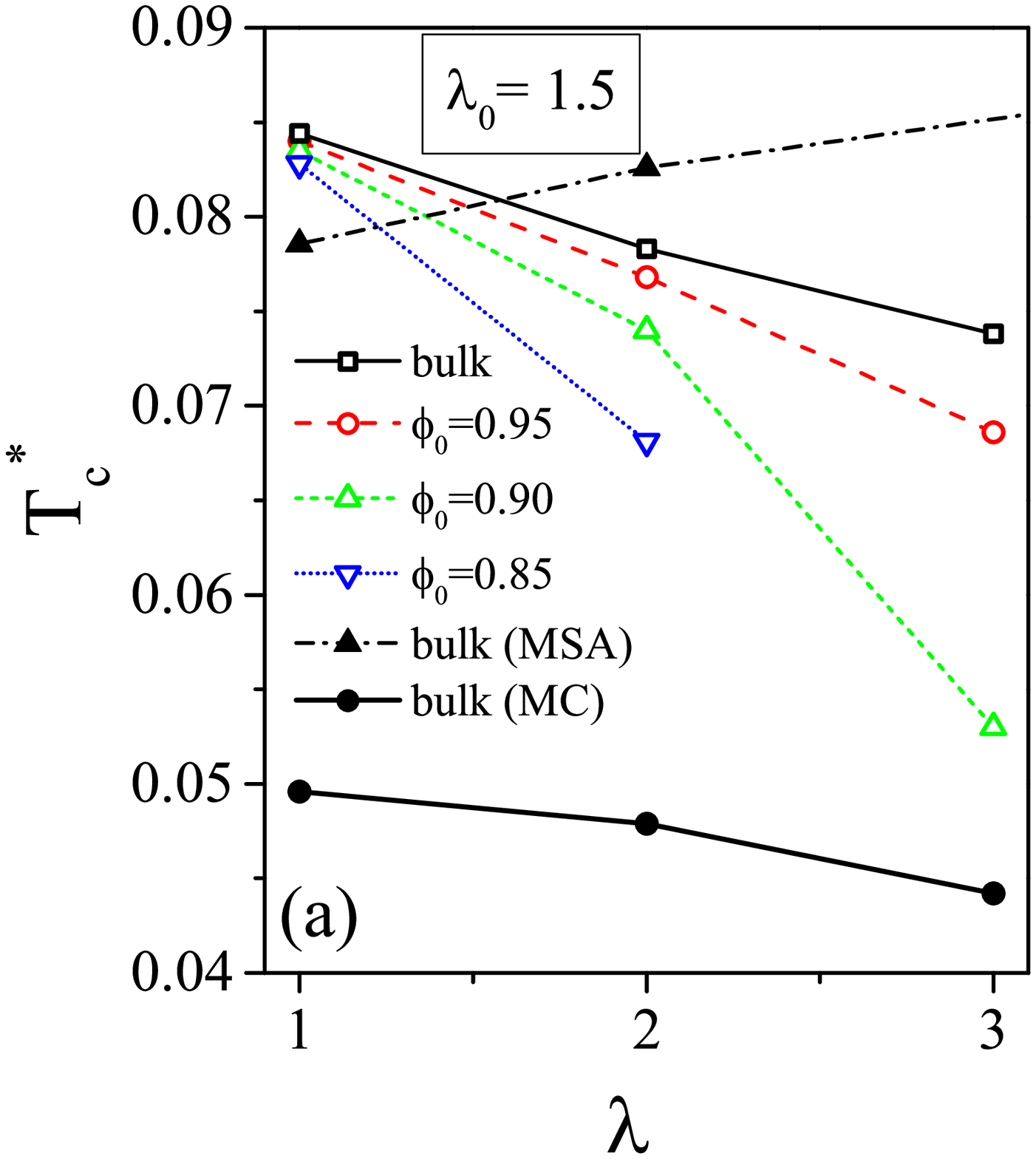}
\includegraphics[clip,width=0.32\textwidth,
angle=0]{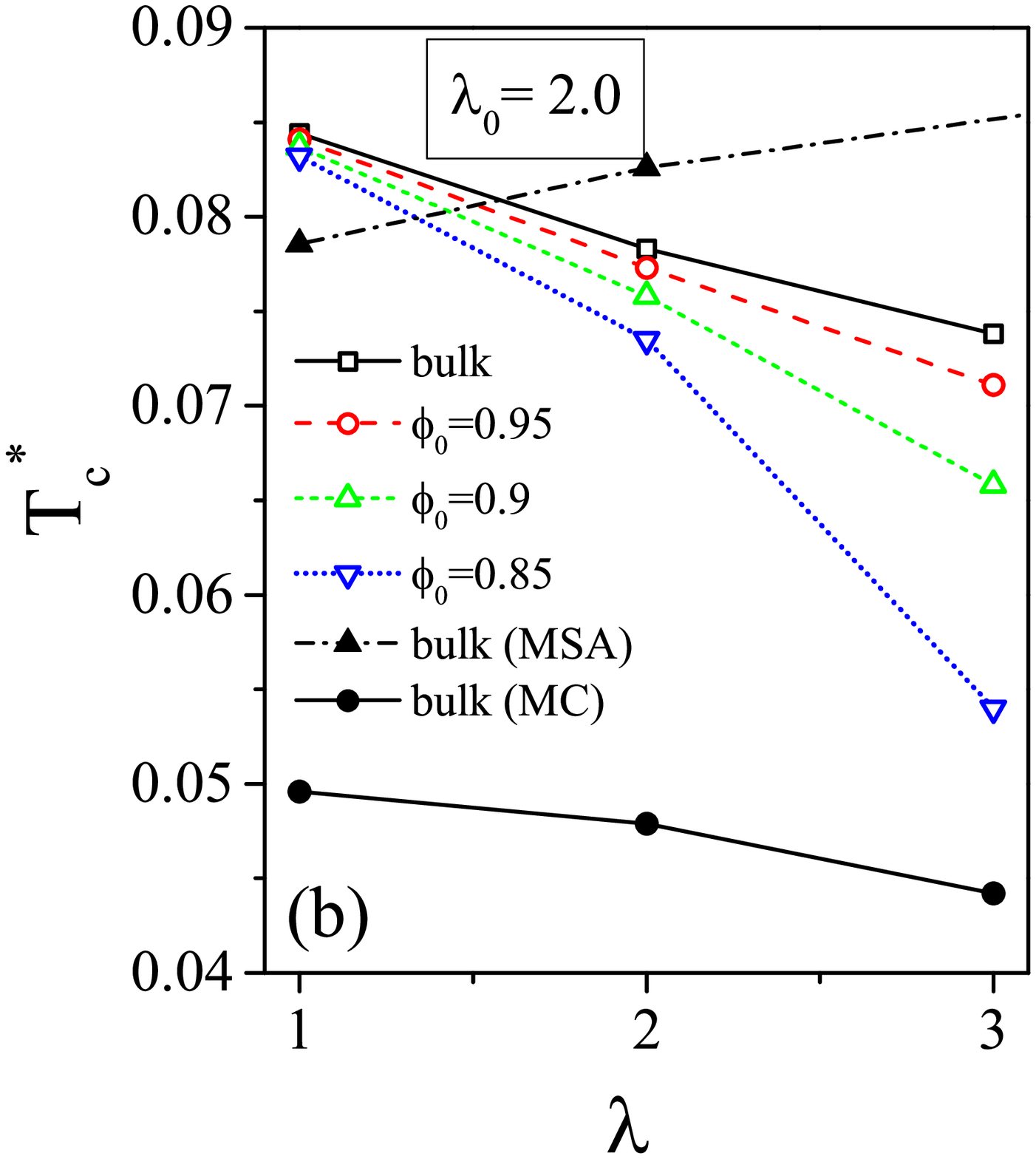}
\includegraphics[clip,width=0.32\textwidth,
angle=0]{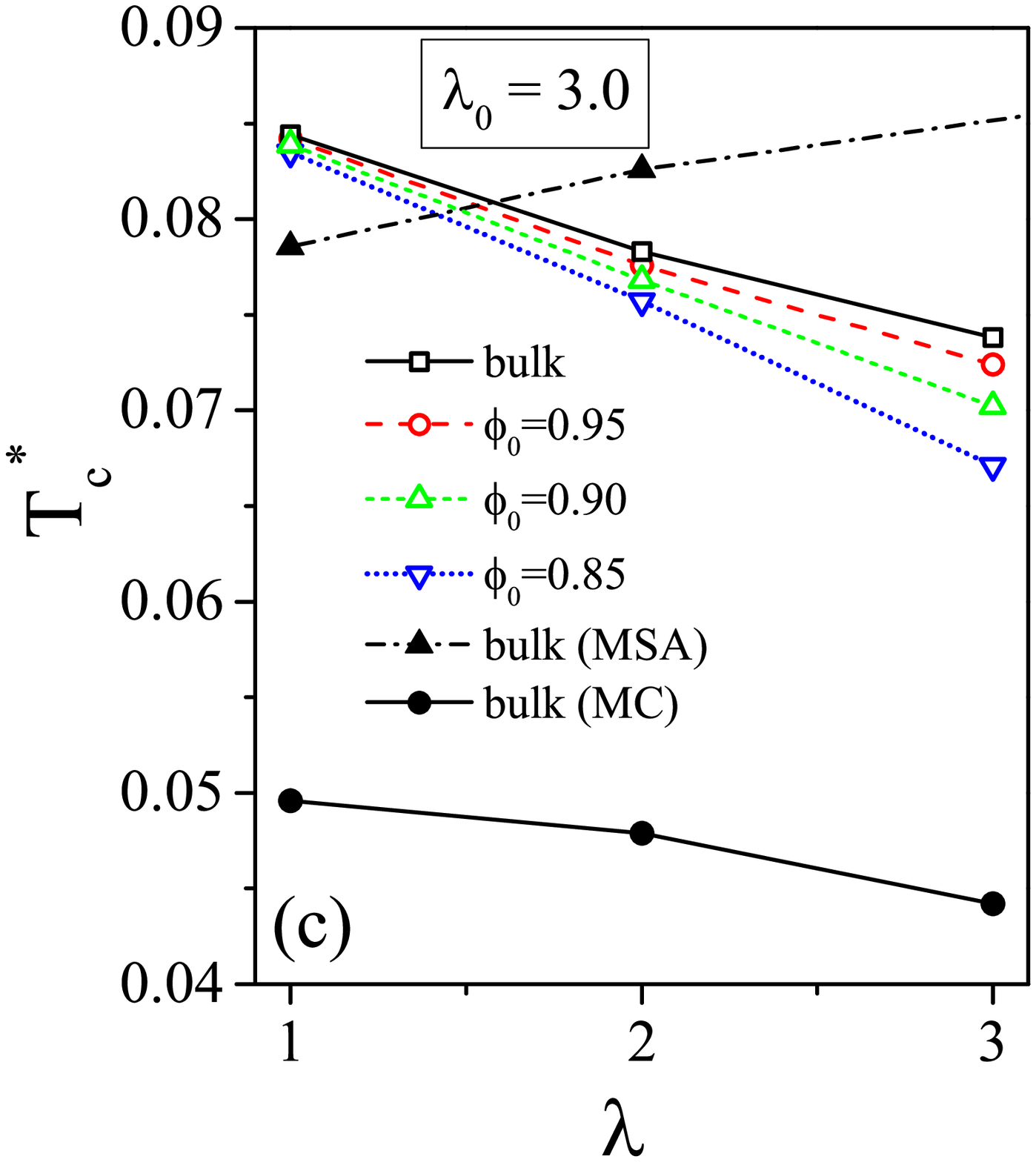}
\caption{\label{fig:Tclamb}
(Color online) Dependence of the critical temperature $T_{c}^{*}$ on the  ion size ratio $\lambda$   at the fixed size ratios $\lambda_{0}=1.5$ (a),  $\lambda_{0}=2.0$ (b), and $\lambda_{0}=3.0$~(c) [see Eq.~(\ref{lambda0})]. In each case, data are for different values
of the matrix porosity as indicated in the legends.
Open symbols denote the results obtained in this study using the CV approach,
filled triangles are  MSA results  \cite{Gonzalez-Tovar}, and  filled circles denote simulation results
\cite{Romero-Enrique:00}.
}
\end{center}
\end{figure}

Similar to the bulk case, the critical temperature $T_{c}^{*}$ of a confined ionic fluid decreases with an increase of
 size asymmetry of ions $\lambda$. The corresponding results are shown in Figs.~\ref{fig:Tclamb}(a)--\ref{fig:Tclamb}(c) for the matrix porosities $\phi_{0}=0.85$, $0.9$ and $0.95$ and for different sizes
of the matrix particles $\lambda_{0}=1.5$, $2.0$ and $3.0$. By comparison, in these figures we  show the critical 
temperatures of a bulk PM fluid obtained from the grand canonical Monte Carlo simulations~\cite{Romero-Enrique:00} and from the calculations performed in the MSA \cite{Gonzalez-Tovar}. It is seen that the dependence of the critical temperature
of a PM fluid on the ion size asymmetry provided within our theoretical approach is in a qualitative agreement with the simulation studies, while the MSA leads to the results with the opposite trend which is considered to be wrong.
For a confined PM fluid, the slopes of the dependence of $T_{c}^{*}$ on $\lambda$ indicate that at lower matrix porosities, the critical temperature decreases faster with $\lambda$. The same effect is noticed when the size of matrix particles is smaller.

\begin{figure}[htb]
\begin{center}
\includegraphics[clip,width=0.32\textwidth,
angle=0]{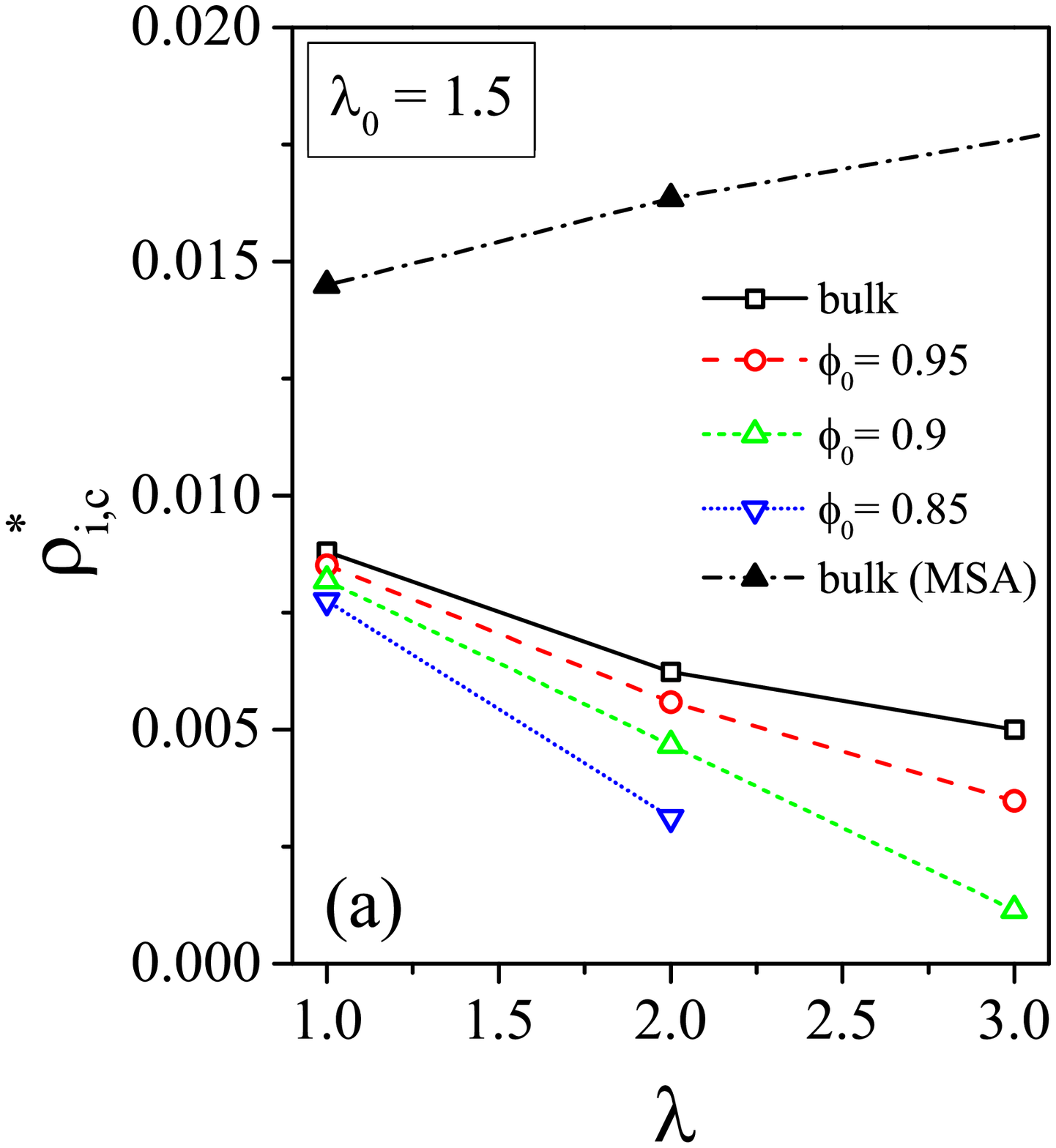}
\includegraphics[clip,width=0.32\textwidth,
angle=0]{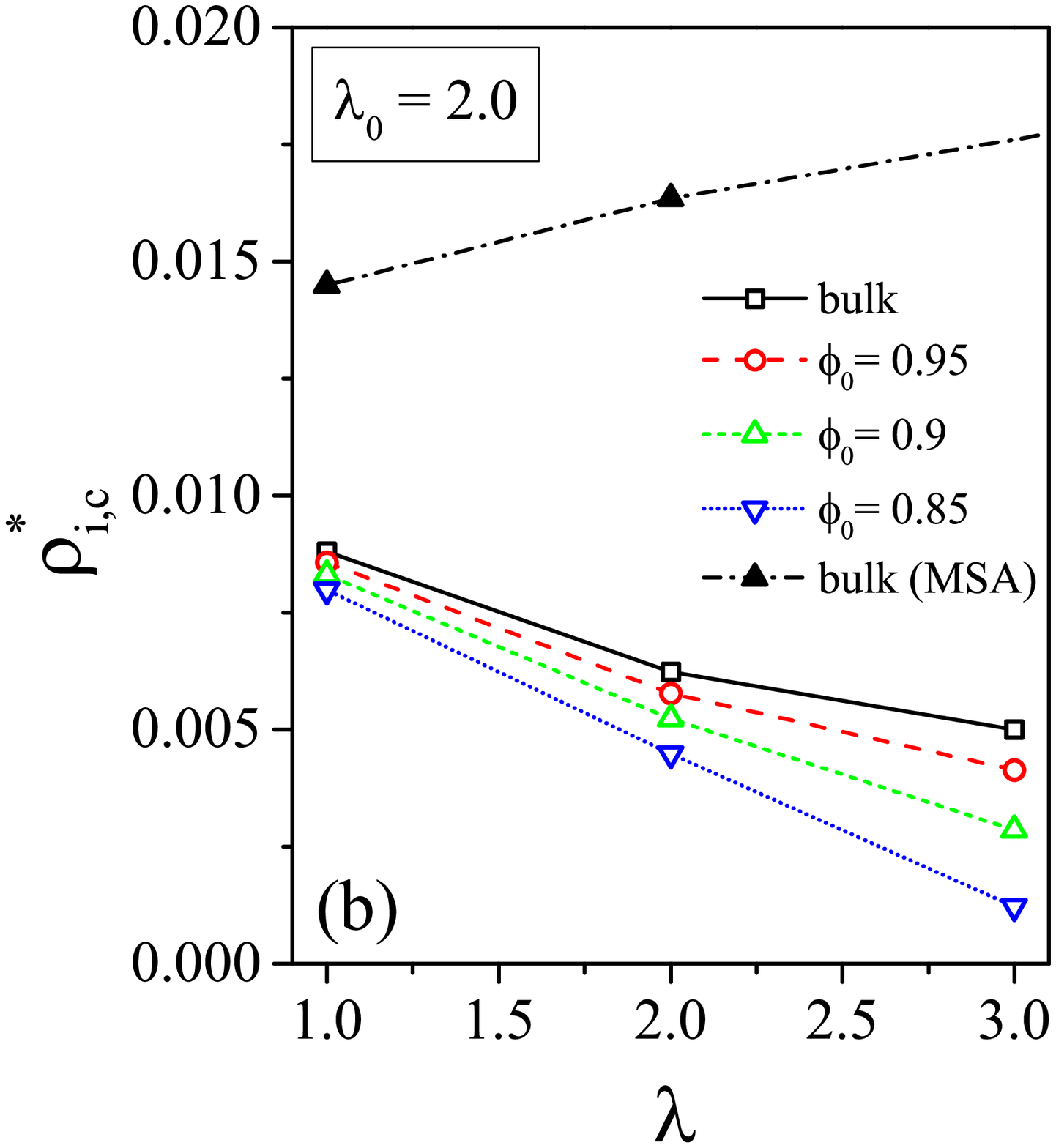}
\includegraphics[clip,width=0.32\textwidth,
angle=0]{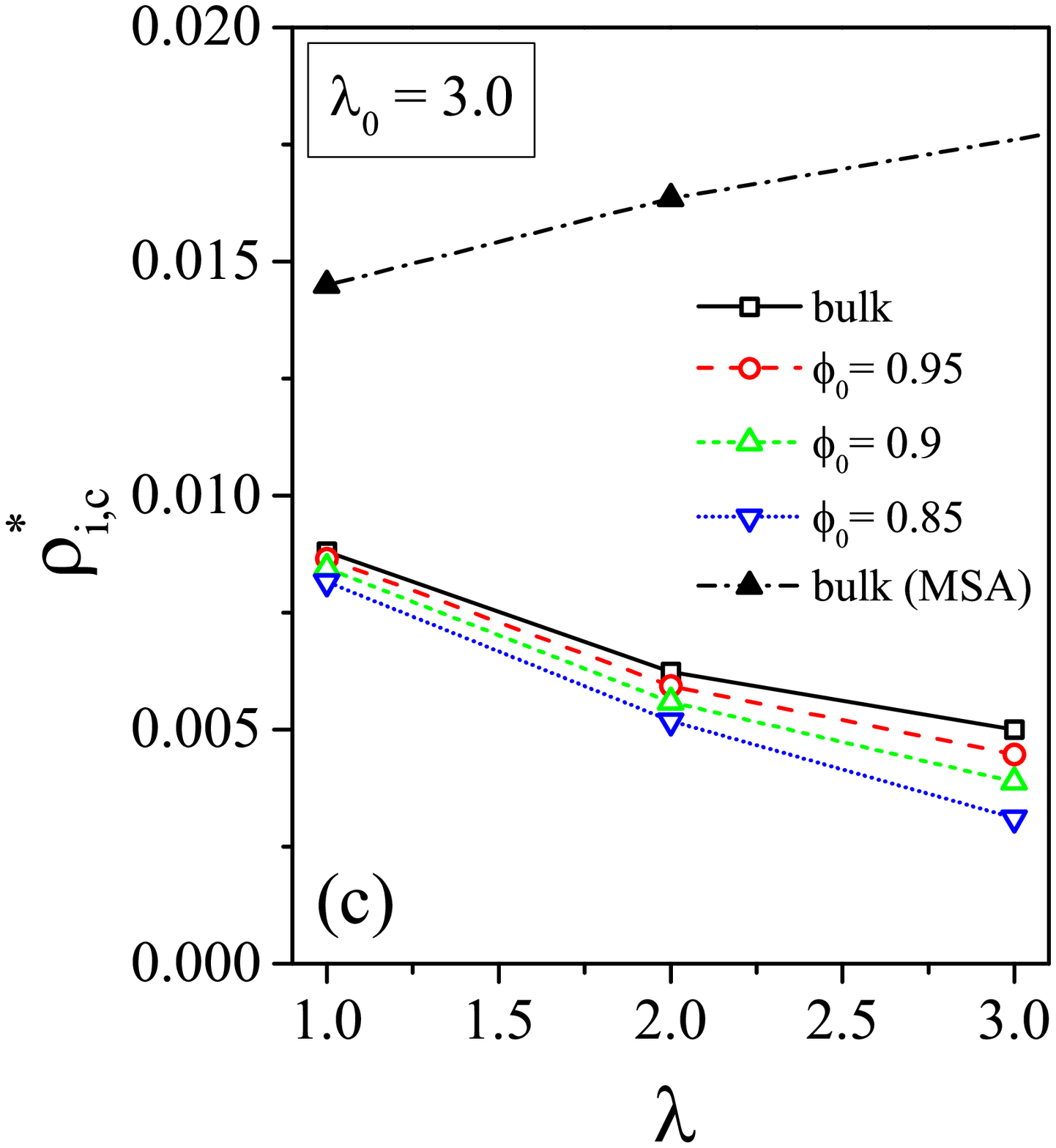} \caption{\label{fig:rhoclamb}
(Color online) Dependence of the critical density $\rho_{i,c}^{*}$ on the  ion size ratio $\lambda$  at the fixed size ratios $\lambda_{0}=1.5$ (a),  $\lambda_{0}=2.0$ (b), and $\lambda_{0}=3.0$~(c) [see Eq.~(\ref{lambda0})].  In each case, data are for different values
of the matrix porosity as indicated in the legends.
Open symbols denote the results obtained in this study using the CV approach and filled triangles are  MSA results  \cite{Gonzalez-Tovar}.
}
\end{center}
\end{figure}

The critical density $\rho_{i,c}^{*}$ of the confined PM fluid depending on $\lambda$ is shown in Figs.~\ref{fig:rhoclamb}(a)--\ref{fig:rhoclamb}(c) for the same matrix porosities and
sizes of matrix particles as in the previous figures. As one can see,  general conclusions
on the behavior of the critical density $\rho_{i,c}^{*}$ qualitatively repeat the conclusions on the behavior of the
critical temperature $T_{c}^{*}$.  It should be noted that  our theoretical predictions  of the trend of
$\rho_{i,c}^{*}$ with $\lambda$
is in a qualitative  agreement with the simulation findings for the  bulk PM  \cite{Romero-Enrique:00}.
On the other hand, the MSA approach again provides  a wrong trend.

\section{Conclusions}
We have studied the vapor-liquid phase equilibrium of an asymmetric binary ionic model confined in a disordered porous medium
formed by a HS matrix. To this end, considering the whole system as a partly-quenched model, we have developed a theoretical 
approach that  enables us to formulate a perturbation theory. The approach is based on the CV method with a reference system. For an asymmetric PM in the bulk state, it allowed us to obtain the correct trends of both the critical temperature and the critical density with size and charge asymmetry. Following the ideas earlier proposed for the bulk PM, we have derived an explicit expression for the relevant chemical potential conjugate to the order parameter which includes
the effects of correlations up to third order. It should be emphasized that the expression takes into account both  
charge and size asymmetry at the same level of  approximation.

In this paper,  the reference system is considered to be  a two-component HS fluid confined in a disordered HS matrix.
The HS fluid is characterized by the  parameter of size asymmetry $\lambda=\sigma_{+}/\sigma_{-}$ while the matrix is
characterized by the diameter
of obstacles $\sigma_{0}$ and different types of matrix porosity, geometrical porosity $\phi_{0}$, and
two probe-particle porosities $\phi_{+}$ and $\phi_{-}$. The description of the reference system has been carried out by using
the recent generalization of the SPT theory for a multicomponent fluid in a multicomponent matrix. Here, we have presented   explicit
expressions for the partial chemical potentials in the approximation that provides the best accuracy against  the simulation results.
Based on these expressions, we have found analytical formulas for the two- and three-body correlation functions of the reference system
in the long-wavelength limit.

Using  an expression for the relevant chemical potential, we have calculated  the vapor-liquid phase diagrams of a monovalent
PM with $\lambda=1$, $2$ and $3$ confined in the HS matrix of different porosities $\phi_0=0.85$, $0.90$ and $0.95$ and
with different size ratios between the matrix obstacles and the negatively charged ions,
$\lambda_{0}=\sigma_{0}/\sigma_{-}=1$, $1.5$, $2$ and $3$. Based on the  phase diagrams, the critical parameters $T_{c}^{*}$
and $\rho_{i,c}^{*}$ of a confined PM fluid have been obtained. It has been shown that both the critical temperature and the critical density lower when the matrix porosity decreases.
On the other hand, at a fixed porosity, the critical parameters $T_{c}^{*}$ and $\rho_{i,c}^{*}$ are higher
in a matrix of large particles than in a matrix of small particles. An increase in the ratio of ion size asymmetry
$\lambda$ leads to the lowering of $T_{c}^{*}$ and $\rho_{i,c}^{*}$, and this trend is essentially strengthened by
the confinement effect at lower porosities, especially when a matrix is composed of the  particles of small sizes.
It should be noted that  variations in the critical parameters $T_{c}^{*}$ and $\rho_{i,c}^{*}$ with $\lambda$ and $\phi_{0}$  confirm our previous results obtained within the framework of the AMSA where,
however, only the case of fixed $\lambda_{0}=2$ is considered. A distinguishing feature of the present approach is the possibility to derive,  without additional assumption such as the presence of ion pairs, an analytical expression for the chemical potential conjugate to the order parameter which provides a qualitatively correct phase behavior of a rather complex system.

Finally,  for the asymmetric PM confined in a disordered porous matrix, we have proposed  an analytical approach which
allows one to make qualitative predictions of the
vapor-liquid phase behavior  depending on size and charge asymmetry of ions
and on  matrix  characteristics such as a geometrical porosity and a diameter of solid obstacles.
We expect that  taking into account of higher-order correlations will lead to
quantitative, but not qualitative, changes to our results.
It should be noted that the present approach can be extended  for
more complex models, e.g., for the models which include  attractive or repulsive ion-matrix interactions in addition to a 
hard-core repulsion.

\acknowledgments
This project has received funding from the European Union's Horizon 2020 research and
innovation programme under the Marie Sk{\l}odowska-Curie grant agreement No 734276, and from the State Fund For
Fundamental Research of Ukraine (project No. F73/26-2017).

\appendix

\section{Grand partition function of a $(2s+1)$-component replicated model in the MF approximation}

The MF part of the grand partition function is of the form:
\begin{equation}
\Xi^{\rm{mf}}=\Xi^{r}[\tilde{\nu}_{0},\tilde{\nu}_{A}^{\alpha}]\exp\left\lbrace\langle N_{0}\rangle_{r}\left[\frac{\beta}{2}
\overline{\rho_{0}}\tilde{u}_{00}^{(p)}(0)
+\sum_{\alpha}\sum_{A}\beta\overline{\rho^{\alpha}_{A}}\tilde{u}_{0A}^{\alpha(p)}(0) \right]\right\rbrace,
\label{Xi_mf}
\end{equation}
where  $\Xi^{r}$ is the grand partition function of a $(2s+1)$-component reference system with the renormalized partial chemical potentials
\begin{eqnarray}
\tilde{\nu}_{0}=\nu_{0}+\frac{\beta}{2V}\sum_{{\mathbf
k}}\tilde{u}_{00}^{(p)}(k)-\overline{\rho_{0}}\beta\tilde{u}_{00}^{(p)}(0)-\sum_{\alpha}\sum_{A}\overline{\rho^{\alpha}_{A}}
\beta\tilde{u}_{0A}^{\alpha(p)}(0),
\label{tilde_nu0} \\
\tilde{\nu}_{A}^{\alpha}=\nu_{A}^{\alpha}+\frac{\beta}{2V}\sum_{{\mathbf
k}}\tilde u_{AA}^{\alpha\alpha(p)}(k)-\overline{\rho_{0}}\beta\tilde{u}_{0A}^{\alpha(p)}(0)-
\sum_{B}\overline{\rho_{B}^{\alpha}}\beta\tilde{u}_{AB}^
{\alpha\alpha(p)}(0),
\label{tilde_nu}
\end{eqnarray}
 $\nu_{0}=\beta\mu_{0}-\ln\Lambda_{0}^{3}$ and $\nu_{A}^{\alpha}=\beta\mu_{A}^{\alpha}-\ln\Lambda_{A}^{3}$
are the dimensionless chemical potentials of the corresponding species ($\Lambda_{0}$ and
$\Lambda_{A}$ are the  de Broglie thermal wavelengths), $\overline{\rho_{0}}=\langle N_{0}\rangle_{r}/V$,
$\overline{\rho^{\alpha}_{A}}=\langle N_{A}^{\alpha}\rangle_{r}/V$, $\langle\ldots\rangle_{r}$
indicates the  average taken over the reference system, $\tilde{u}_{00}^{(p)}(k)$, $\tilde{u}_{0A}^{\alpha(p)}(k)$ and $\tilde{u}_{AB}^{\alpha\alpha(p)}(k)$ are the
Fourier transforms of the perturbative parts of the corresponding interaction potentials.

\section{Expressions for ${\mathfrak{M}}_{\alpha_{1}\alpha_{2}\alpha_{3}}(0,0)$   }
For a two-component system, the expressions for the third-order cumulants in the long-wavelength limit are as follows:
\begin{eqnarray}
{\mathfrak{M}}_{+++}(0,0)&=&\rho_{+}\left\{S_{++}(0)\left[S_{++}(0)+\eta_{+}\left(\frac{\partial S_{++}(0)}
{\partial\eta_{+}}\right)_{\eta_{-}}\right]\right. \nonumber \\
&&
\left.
+\sqrt{\frac{\rho_{+}}{\rho_{-}}}S_{+-}(0)\,\eta_{-}\left(\frac{\partial S_{++}(0)}{\partial\eta_{-}}\right)_{\eta_{+}}\right\}
\label{M_+++}
\end{eqnarray}
\begin{eqnarray}
{\mathfrak{M}}_{++-}(0,0)&=&\sqrt{\rho_{+}\rho_{-}}\left\lbrace S_{+-}(0)\left[ S_{++}(0)+\eta_{+}\left(\frac{\partial S_{++}(0)}
{\partial\eta_{+}}\right)_{\eta_{-}}\right]\right. \nonumber \\
&&
\left.
+\sqrt{\frac{\rho_{+}}{\rho_{-}}}S_{--}(0)\,\eta_{-}\left(\frac{\partial S_{++}(0)}{\partial\eta_{-}}\right)_{\eta_{+}}
\right\rbrace.
\label{M_++-}
\end{eqnarray}
The expressions for ${\mathfrak{M}}_{---}(0,0)$ and ${\mathfrak{M}}_{+--}(0,0)$ can be obtained replacing indices ``$+$'' by
indices ``$-$'' and vice versa. The same formulas are valid for the connected parts of the corresponding quantities.

\end{document}